\documentclass[11pt]{article}

\usepackage{amsmath,amssymb,amsthm,mathtools,bm}
\usepackage{geometry}
\usepackage[authoryear,round]{natbib}
\usepackage{url}
\usepackage{enumitem}
\usepackage{booktabs}
\usepackage{microtype}
\usepackage{lineno}

\newtheorem{theorem}{Theorem}[section]
\newtheorem{proposition}[theorem]{Proposition}
\newtheorem{lemma}[theorem]{Lemma}
\newtheorem{corollary}[theorem]{Corollary}
\newtheoremstyle{boldremark}{\topsep}{\topsep}{\normalfont}{}{\bfseries}{.}{.5em}{}
\theoremstyle{boldremark}
\newtheorem{remark}[theorem]{Remark}
\makeatletter
\renewenvironment{proof}[1][\proofname]{\par\pushQED{\qed}\normalfont
  \topsep6\p@\@plus6\p@\relax\trivlist
  \item[\hskip\labelsep\bfseries\itshape #1\@addpunct{.}]\ignorespaces}
  {\popQED\endtrivlist\@endpefalse}
\makeatother

\newcommand{\R}{\mathbb R}

\newcommand{\Ep}{\mathbb E}
\newcommand{\ind}{\mathbf 1}
\newcommand{\thet}{\boldsymbol\theta}
\newcommand{\thetao}{\boldsymbol\theta_0}
\newcommand{\Xv}{\mathbf X}
\newcommand{\xv}{\mathbf x}

\newcommand{\Op}{O_p}

\renewcommand{\baselinestretch}{1.2}

\title{Strong Consistency and Asymptotic Normality of the Maximum Likelihood
Estimator in $r$-Largest Generalized Extreme-Value Models}
\author{Jihong Park$^{1}$, Yire Shin$^{2,*}$, Jeong-Soo Park$^{2}$\\[1ex]
\small $^1$Department of Mathematics and Statistics, Chonnam National University, Gwangju 61186, Korea\\
\small $^2$Department of Statistics, Chonnam National University, Gwangju 61186, Korea\\
\small $^*$Corresponding author: shinyire87@gmail.com}
\date{}

\begin{document}
\maketitle

\begin{abstract}
We study maximum likelihood estimation in the three-parameter $r$-largest
generalized extreme-value ($r$GEV) model with a fixed number $r$ of upper order
statistics per block. Because the support depends on all three parameters,
the classical regular theory does not apply. We show that the maximum
likelihood estimator is strongly consistent on compact parameter sets
contained in $k<1$, where $k=-\gamma$ denotes the shape parameter. For $k_0<1/2$,
the range in which the model is differentiable in quadratic mean, the
estimator is root-$n$ consistent, asymptotically normal and asymptotically
efficient. The proof verifies the conditions of the moving-support likelihood
theorem of B\"ucher and Segers for the ordered $r$-dimensional likelihood; the
main new ingredients are a sequential modified likelihood criterion and a
support-loss bound showing that the regularity threshold $k_0<1/2$ does not
depend on $r$. In simulated heavy-tailed cases with only $50$ blocks, using
five to seven order statistics instead of the block maximum reduces the
relative bias of the 100-year return level from $6$--$13\%$ to below $5\%$.
\end{abstract}

\noindent{\bf Keywords}: Asymptotic efficiency; asymptotic normality;
maximum likelihood; moving support; strong consistency;
$r$ largest order statistics.

\section{Introduction}
\label{sec:introduction}

Likelihood inference for the three-parameter generalized extreme-value (GEV)
distribution is not a routine application of regular theory, because the
support depends on the parameters. \citet{Smith1985} showed for families with
moving endpoints that the local behavior of the density at the boundary
determines whether the classical root-$n$ normal theory survives, and
\citet{ChengTraylor1995} surveyed non-regular maximum likelihood problems more
broadly. A rigorous modern theory of maximum likelihood in extreme-value
models has since been built in a series of works. In the peaks-over-threshold
setting, \citet{DreesFerreiradeHaan2004} established asymptotic normality of the
maximum likelihood estimator of the extreme value index, and
\citet{Zhou2009} settled its existence and consistency. For the GEV family,
\citet{buchersegers2017maximum} proved strong consistency of the maximum
likelihood estimator for $\gamma_0>-1$, and gave a general theorem for
asymptotic linearity of maximum likelihood estimators in models with
parameter-dependent support, which they applied for $\gamma_0>-1/2$. In the
convention $k=-\gamma$ used here, these ranges are $k_0<1$ and $k_0<1/2$. This
theorem is the backbone of the present paper: it isolates precisely which
properties of a moving-support model have to be checked. A complementary line
of work treats the GEV distribution as an approximation to block maxima rather
than as the exact sampling model: \citet{dombry2015existence} established
existence and consistency of the maximum likelihood estimator in the
block-maxima framework,
\citet{dombry2019maximum} established asymptotic normality of the maximum
likelihood estimator under a second-order condition and quantified the bias of
the finite-block approximation, and \citet{BucherSegers2018Frechet} treated
block maxima extracted from stationary time series, including sliding blocks.

The $r$GEV model, which uses the $r$ largest observations of each block
\citep{smith1986extreme,tawn1988extreme}, is widely used for environmental
extremes \citep{soares2004application,an2007r,wang2008downscaling,hamdi2014extreme,muhammed2017variations,e2020change}.
\citet{dupuis1997extreme} proposed a robust estimation approach for the $r$
largest annual events. The model has also been extended in several directions
\citep{shin2023modeling,ShinPark2024Serra,ShinPark2025SciRep,Busaba2025CSAM,Busababodhin2026,silva2022dynamic}.
It is implemented in the R packages \texttt{ismev} \citep{ismev2025}, which
accompanies \citet{coles2001introduction}, \texttt{eva} \citep{bader2020package}
and \texttt{evmr} \citep{ShinParkRpackageevmr}.

The choice of $r$ is studied by \citet{bader2017automated} and
\citet{ShinParkPark2026}. In all of these works the standard errors of the
estimates and of derived quantities such as return levels are computed from
the inverse observed information; that is, the usual normal approximation for
the maximum likelihood estimator is taken for granted. The selection procedure
of \citet{ShinParkPark2026}, for instance, imposes this approximation for each
candidate $r$ as a regularity assumption. For the exact three-parameter $r$GEV
likelihood with moving support, however, it has not been justified. \citet{BucherHaufs2026} note that theory
beyond the block maximum has remained limited; in a time-series setting they
develop a careful theory for blockwise top-two order statistics and show that a
naive independence likelihood can be inconsistent under serial dependence. The
present paper is complementary: it treats any fixed $r$, but only for
independent blocks with the exact $r$GEV law.

In a companion paper \citep{ParkShinPark2026A} we showed that the $r$GEV family
with fixed $r$ is differentiable in quadratic mean if and only if $k_0<1/2$,
that its Fisher information $I_r(\thetao)$ is positive definite in this range
and equals the negative expected Hessian, and that the model is then locally
asymptotically normal. The present paper turns from the model to the
estimator. For independent $r$GEV vectors $\Xv_1,\ldots,\Xv_n$ with $r$ fixed,
we show that the maximum likelihood estimator $\widehat\thet_n$ is strongly
consistent on compact parameter sets contained in $k<1$
(Theorem~\ref{thm:consistency}), a range wider than the regular one, and that
for $k_0<1/2$ it is root-$n$ consistent (Proposition~\ref{prop:rootn}) with
\[
\sqrt n(\widehat\thet_n-\thetao)
=
I_r(\thetao)^{-1}
\frac1{\sqrt n}\sum_{i=1}^n
\dot{\boldsymbol\ell}_{r,\thetao}(\Xv_i)
+o_p(1).
\]
Hence it is asymptotically normal and efficient
(Theorem~\ref{thm:asymp-linear} and Corollary~\ref{cor:efficient}), and
return-level estimators are asymptotically normal with the delta-method
variance (Corollary~\ref{cor:return-level}).

The asymptotic normality is obtained from
\citet[Proposition~2.1]{buchersegers2017maximum}, whose four conditions are
differentiability in quadratic mean with nonsingular information, a negligible
probability of leaving the common support of a shrinking neighborhood, a
Lipschitz bound for the log-likelihood on that support, and the root-$n$ rate.
The first is the companion result; the other three need new arguments for the
ordered $r$-dimensional likelihood. The main difficulty lies in the rate. The
modified-criterion method of \citet{buchersegers2017maximum} replaces the
log-likelihood ratio by a bounded-below criterion built from a power of the
density, but for the joint density of the $r$ largest order statistics a single
power works only for $k_0<1/3$ (Appendix~\ref{app:rate}): the largest
coordinate of an observation vector can be close to the moving endpoint while
its smallest lies far in the lower tail. For $0<k_0<1/2$ we therefore apply the
criterion separately to the marginal factor of $X^{(1)}$ and to the conditional
factors of $X^{(s)}$ given $X^{(s-1)}$, and for $k_0\le0$ we reduce the drift of
a joint criterion to that of the first marginal by the data-processing
inequality. For the support conditions, the arrival representation shows that
only the leading maximum (for $k_0>0$) or the smallest retained order statistic
(for $k_0<0$) can leave the common support, which gives a support-loss
probability $o(\varepsilon^2)$ and an $L^2$ envelope exactly when $k_0<1/2$.
The regularity threshold is therefore the same for every fixed $r$ as for the
block-maximum GEV model.

We assume independent blocks without further serial dependence or clustering,
which in applications corresponds to filtering or declustering
\citep{tawn1988extreme}. This assumption can be checked with goodness-of-fit
tests based on the conditional distribution functions of successive order
statistics, that is, on the Rosenblatt transform of the $r$GEV vector, which
\citet{ShinParkPark2026} proposed for selecting $r$; a similar construction is
used by \citet{heinonen7451430dependence} to detect serial dependence.
Inference based on the maximum likelihood estimator (observed information,
likelihood ratio tests and the parametric bootstrap) is the subject of ongoing
work; for bootstrap consistency in the block-maxima setting see
\citet{BucherStaud2026}.

Section~\ref{sec:model} recalls the model and the results of
\citet{ParkShinPark2026A} used here, Section~\ref{sec:consistency} proves
strong consistency, Section~\ref{sec:results} establishes the rate, asymptotic
normality and efficiency, Section~\ref{sec:simulation} examines finite-sample
behavior, and Section~\ref{sec:discussion} discusses the results. The proofs
are in the appendix.

\section{Model and results from the companion paper}
\label{sec:model}

\subsection{The $r$GEV model with fixed $r$}

Throughout, $r$ is a fixed finite positive integer, and $n$, the number of
independent observation vectors
$\Xv_i=(X_i^{(1)},\ldots,X_i^{(r)})$, tends to infinity. We treat the $r$GEV law
as the exact model for these vectors; regimes with $r=r_n\to\infty$, with a
data-selected $\widehat r_n$, or with finite-block approximation error are not
covered.

Let $\thet=(\mu,\sigma,k)$ with $\sigma>0$, and put
$w_{\thet}(x)=1-k(x-\mu)/\sigma$ and, for $k\ne0$,
\[
T_{\thet}(x)=w_{\thet}(x)^{1/k},
\]
with the continuous limit $T_{(\mu,\sigma,0)}(x)=\exp\{-(x-\mu)/\sigma\}$. The GEV
distribution function is $F_{\thet}(x)=\exp\{-T_{\thet}(x)\}$, and $f_{\thet}$ denotes
its density.

\begin{remark}[Shape-parameter convention]
\label{rem:shape-convention}
This is the sign convention of \citet{prescott1980maximum},
\citet{tawn1988extreme} and \citet{hosking1997regional}, which is common in
hydrology. Much of the modern literature, e.g.\
\citet{coles2001introduction}, \citet{dombry2019maximum} and
\citet{BucherZhou2021}, writes the GEV law as
$\exp[-\{1+\gamma(x-\mu)/\sigma\}^{-1/\gamma}]$ with shape $\gamma$ or $\xi$. The two
conventions are related by
\begin{equation}
\gamma=\xi=-k.
\label{eq:shape-sign-conversion}
\end{equation}
Thus $\gamma>-1/2$ corresponds to $k<1/2$, $\gamma>-1$ to $k<1$, and $k>0$ to a
finite upper endpoint $\omega(\thet)=\mu+\sigma/k$.
\end{remark}

For fixed $r\ge1$, the joint density of the $r$ largest order statistics is
\citep[Section~3.5]{coles2001introduction}
\begin{equation}
 p_{r,\thet}(x_1,\ldots,x_r)
 =\sigma^{-r}\exp\{-T_{\thet}(x_r)\}
 \prod_{j=1}^r T_{\thet}(x_j)^{1-k},
 \label{eq:rgev-density}
\end{equation}
on the support
\[
S_{r,\thet}
=\{\xv=(x_1,\ldots,x_r)\in\mathbb R^r:
x_1>\cdots>x_r,\;
w_{\thet}(x_j)>0,\ j=1,\ldots,r\}.
\]
We write $\ell_{r,\thet}=\log p_{r,\thet}$, with $\ell_{r,\thet}=-\infty$ outside
$S_{r,\thet}$, $P_{r,\thet}$ for the corresponding law, and $\Ep_{\thet}$ for
expectation under it.

At the true parameter $\thetao=(\mu_0,\sigma_0,k_0)$, the variables
$T_j=T_{\thetao}(X^{(j)})$ are the first $r$ arrival times
$0<T_1<\cdots<T_r$ of a unit-rate Poisson process
\citep{Weissman1978,Resnick1987}:
\begin{equation}
T_1,\ T_2-T_1,\ldots,T_r-T_{r-1}
\stackrel{\rm ind}{\sim}\operatorname{Exp}(1),
\label{eq:arrival}
\end{equation}
so that $T_j\sim\operatorname{Gamma}(j,1)$. This law does not depend on $\thetao$;
equivalently, the copula of the $r$ largest order statistics is parameter free
\citep{heinonen7451430dependence}.
In terms of $T_j=T_{\thet}(x_j)$ the log-density is
\begin{equation}
\ell_{r,\thet}
=-r\log\sigma-T_r+(1-k)\sum_{j=1}^r\log T_j.
\label{eq:loglik-T}
\end{equation}
Let $D_{\thet}(x)=\nabla_{\thet}\log T_{\thet}(x)$. For $k\ne0$,
\begin{equation}
D_{\thet}(x)=
\begin{pmatrix}
T_{\thet}(x)^{-k}/\sigma\\[1mm]
\{T_{\thet}(x)^{-k}-1\}/(k\sigma)\\[1mm]
-\log T_{\thet}(x)/k-
\{T_{\thet}(x)^{-k}-1\}/k^2
\end{pmatrix},
\label{eq:D}
\end{equation}
with continuous limits at $k=0$. Writing $D_j=D_{\thet}(x_j)$, the score is
\begin{equation}
\dot{\boldsymbol\ell}_{r,\thet}
=-\frac r\sigma\bm e_\sigma
-T_rD_r
+(1-k)\sum_{j=1}^rD_j
-\bm e_k\sum_{j=1}^r\log T_j.
\label{eq:joint-score}
\end{equation}
Finally, the joint density factorizes sequentially as
\begin{equation}
p_{r,\thet}(x_1,\ldots,x_r)
=
f_{\thet}(x_1)
\prod_{s=2}^r
q_{s,\thet}(x_s\mid x_{s-1}),
\qquad
q_{s,\thet}(y\mid x)
=
\frac{f_{\thet}(y)}{F_{\thet}(x)}
\ind(y<x),
\label{eq:sequential-factorization}
\end{equation}
where $q_{s,\thet}(\cdot\mid x)$ is the conditional density of $X^{(s)}$ given
$X^{(s-1)}=x$; by the Markov property of the order statistics, it is also the
conditional density of $X^{(s)}$ given $X^{(1)},\ldots,X^{(s-1)}$.

\subsection{Results from the companion paper}

We use the following results of \citet{ParkShinPark2026A}; their proofs are not
repeated here. Let $s_{\thet}=p_{r,\thet}^{1/2}$, extended by zero outside
$S_{r,\thet}$, let $\Theta_{1/2}=\{(\mu,\sigma,k):\sigma>0,\ k<1/2\}$, and let
\[
I_r(\thet)
=\Ep_{\thet}
\bigl[
\dot{\boldsymbol\ell}_{r,\thet}
\dot{\boldsymbol\ell}_{r,\thet}^{T}
\bigr].
\]

\begin{theorem}[\citealp{ParkShinPark2026A}, Theorem~3.1]
\label{thm:dqm}
The $r$GEV family with fixed $r$ is differentiable in quadratic mean at
$\thetao$ if and only if $k_0<1/2$. When $k_0<1/2$, the quadratic-mean score is
$\dot{\boldsymbol\ell}_{r,\thetao}$ on the support and zero outside it.
\end{theorem}

\begin{proposition}[\citealp{ParkShinPark2026A}, Proposition~3.2]
\label{prop:info-pd}
If $k_0<1/2$, then $I_r(\thetao)\succeq I_1(\thetao)>0$; in particular
$I_r(\thetao)$ is positive definite. Moreover,
\[
I_r(\thetao)=I_1(\thetao)+\sum_{s=2}^r\Ep_{\thetao}(\bm c_s\bm c_s^T),
\]
where $\bm c_s$ is the score at $\thetao$ of the conditional factor
$q_{s,\thet}$ in \eqref{eq:sequential-factorization}.
\end{proposition}

\begin{theorem}[\citealp{ParkShinPark2026A}, Theorem~3.3]
\label{thm:info-identity}
If $k_0<1/2$, the Hessian $\ddot{\boldsymbol\ell}_{r,\thetao}$ is
$P_{\thetao}$-integrable and
$\Ep_{\thetao}\ddot{\boldsymbol\ell}_{r,\thetao}=-I_r(\thetao)$.
\end{theorem}

\begin{corollary}[\citealp{ParkShinPark2026A}, Corollary~3.4]
\label{cor:lan}
Let $k_0<1/2$.
\begin{enumerate}[label=(\roman*)]
\item The model is locally asymptotically normal at $\thetao$ with information
$I_r(\thetao)$; in particular, $P^n_{\thetao+\bm h_n/\sqrt n}$ and $P^n_{\thetao}$ are
mutually contiguous for every convergent sequence $\bm h_n$, where $P^n_{\thet}$
denotes the joint law of $\Xv_1,\ldots,\Xv_n$ under $\thet$.
\item For every $\psi:\Theta_{1/2}\to\R^m$ differentiable at $\thetao$, the
local asymptotic minimax risk for estimating $\psi(\thetao)$ under bowl-shaped
loss is bounded below by the risk of
$N_m\{0,\dot\psi_{\thetao}I_r(\thetao)^{-1}\dot\psi_{\thetao}^T\}$.
\end{enumerate}
\end{corollary}

The next two auxiliary results are used in the rate proof.

\begin{lemma}[\citealp{ParkShinPark2026A}, proof of Lemma~A.1]
\label{lem:D-bound}
Let $L=\log t$. For every $\sigma>0$ and real $k$,
$D_{\thet}$, written as a function of $t=T_{\thet}(x)$, has the integral
representation
\[
D_\mu=\sigma^{-1}e^{-kL},\qquad
D_\sigma=-\frac{L}{\sigma}\int_0^1e^{-vkL}\,dv,\qquad
D_k=-L^2\int_0^1(1-v)e^{-vkL}\,dv.
\]
Consequently, if $c\le\sigma\le C$ and $-\bar k\le k\le k_+$, then
\begin{equation}
\|D_{\thet}(t)\|
\le
C_1\max(1,t^{-k})\{1+|\log t|\}^2
\le
C_1\{1+t^{-k_+}+t^{\bar k}\}\{1+|\log t|\}^2,
\label{eq:D-uniform}
\end{equation}
with $C_1$ depending only on $c$ and $C$.
\end{lemma}

The first bound follows from $e^{-vkL}=t^{-vk}\le\max(1,t^{-k})$ for $0\le v\le1$;
it contains no factor $1/k$, so it holds uniformly across $k=0$. To handle the
moving support we use the following fact. For fixed real numbers
$x_1,\ldots,x_m$ put $\pi_j(\thet)=\sigma-k(x_j-\mu)$, so that $w_{\thet}(x_j)>0$ if
and only if $\pi_j(\thet)>0$.

\begin{lemma}[Zero extension; \citealp{ParkShinPark2026A}, Lemma~5.1]
\label{lem:zero-extension}
Let $V\subset\{\thet:\sigma>0\}$ be open and convex and let
$O=\{\thet\in V:\pi_j(\thet)>0,\ j=1,\ldots,m\}$. Let $g:V\to\R$ and
$c\in\R$ satisfy:
\begin{enumerate}[label=(\roman*)]
\item $g$ is continuously differentiable on $O$;\qquad (ii) $g\equiv c$ on $V\setminus O$;
\item[(iii)] for every $\thet^*\in V\cap\partial O$, $g(\thet)\to c$ and
$\nabla g(\thet)\to0$ as $\thet\to\thet^*$ with $\thet\in O$.
\end{enumerate}
Then $g$ is continuously differentiable on $V$, with $\nabla g=0$ on
$V\setminus O$, and for all $\thet_1,\thet_2\in V$,
\begin{equation}
|g(\thet_1)-g(\thet_2)|
\le\sup_{\vartheta\in[\thet_1,\thet_2]\cap O}\|\nabla g(\vartheta)\|\,
\|\thet_1-\thet_2\|,
\label{eq:zero-extension-mvt}
\end{equation}
where the supremum over the empty set is zero.
\end{lemma}

Applied to the square-root density, Lemma~\ref{lem:zero-extension} gives the
following.

\begin{lemma}[\citealp{ParkShinPark2026A}, Lemma~A.2]
\label{lem:sqrt-smooth}
Let $V$ be an open convex set whose closure lies in $\{(\mu,\sigma,k):k<1/3\}$
and on which $\sigma$ is bounded away from zero and infinity. Then, for every
$\xv\in\R^r$, the map $\thet\mapsto s_{\thet}(\xv)$ is continuously
differentiable on $V$, with
$\nabla s_{\thet}=\tfrac12s_{\thet}\dot{\boldsymbol\ell}_{r,\thet}$ on
$S_{r,\thet}$ and $\nabla s_{\thet}=0$ off it.
\end{lemma}

\section{Strong consistency}
\label{sec:consistency}

Consistency does not require differentiability in quadratic mean, and it holds
in a wider range of shape parameters than the asymptotic normality of
Section~\ref{sec:results}. Let $P_n$ denote the empirical measure of the
independent blocks $\Xv_1,\ldots,\Xv_n$, so that
$P_nf=n^{-1}\sum_{i=1}^nf(\Xv_i)$ for a measurable function $f$, and let
\[
\Theta_1=\{(\mu,\sigma,k):\sigma>0,\ k<1\}.
\]

\begin{theorem}[Strong consistency]
\label{thm:consistency}
Let $r\ge1$ be fixed and let $\Xv_1,\Xv_2,\ldots$ be i.i.d.\ from the $r$GEV
model with true parameter $\thetao=(\mu_0,\sigma_0,k_0)$, where $\sigma_0>0$ and
$k_0<1$. Let $K$ be a compact subset of $\Theta_1$ with $\thetao$ in its
interior. Then, for every $n$ and every sample, the set
$\arg\max_{\thet\in K}P_n\ell_{r,\thet}$ is nonempty, and there is an event of
probability one on which every sequence of maximizers
$\widehat\thet_n\in\arg\max_{\thet\in K}P_n\ell_{r,\thet}$ satisfies
$\widehat\thet_n\to\thetao$. In particular, every measurable maximizer is
strongly consistent.
\end{theorem}

By Remark~\ref{rem:shape-convention}, the condition
$k_0<1$ is exactly the condition
$\gamma_0>-1$ of \citet[Proposition~3.1]{buchersegers2017maximum}.
The proof, in Appendix~\ref{app:consistency}, is a Wald-type argument
\citep[cf.][Theorem~5.14]{vanderVaart1998}, adapted as in
\citet{dombry2015existence} to log-likelihoods that take the value $-\infty$. Two
features of the fixed-$r$ likelihood make it work. First, for $k<1$ the
log-likelihood is bounded above uniformly in $\thet\in K$ and in the observation,
because \eqref{eq:loglik-T} is at most $-r\log\sigma-T_r+r(1-k)\log T_r$. Second,
$\thet\mapsto\ell_{r,\thet}(\xv)$ is upper semicontinuous for every $\xv$: as a
parameter path approaches the boundary of the support, the log-density tends to
$-\infty$. Identifiability is inherited from the first marginal, which is the
ordinary GEV distribution. The restriction to $k<1$ is natural: for $k>1$ the
density is unbounded near the upper endpoint, so that the likelihood is unbounded
if such shapes are allowed, as for the GEV model \citep[Sections~1 and~7]{Smith1985}.

\section{Root-$n$ rate and asymptotic normality}
\label{sec:results}

\begin{proposition}[Root-$n$ rate]
\label{prop:rootn}
Let $r\ge1$ be fixed, let $k_0<1/2$, and let $K$ be a compact subset of
$\Theta_1$ with $\thetao$ in its interior. If $\widehat\thet_n$ maximizes the
$r$GEV likelihood over $K$, then
\[
\sqrt n(\widehat\thet_n-\thetao)=\Op(1).
\]
\end{proposition}

For $\varepsilon>0$, define
\[
U_\varepsilon(\thetao)
=\{\thet:\|\thet-\thetao\|_\infty<\varepsilon\}.
\]
The common support over this parameter neighborhood is
\begin{equation}
\overline S_r(\varepsilon)
=\bigcap_{\thet\in U_\varepsilon(\thetao)}S_{r,\thet}.
\label{eq:common-support}
\end{equation}

\begin{lemma}[Negligible loss of support]
\label{lem:support-loss}
If $k_0<1/2$, then
\[
P_{\thetao}
\{S_{r,\thetao}\setminus\overline S_r(\varepsilon)\}
=o(\varepsilon^2),
\qquad \varepsilon\downarrow0.
\]
\end{lemma}

\begin{lemma}[Common-support likelihood Lipschitz condition]
\label{lem:loglik-lipschitz}
Let $k_0<1/2$ and let $\|\cdot\|_\nu$ be either the maximum norm
($\nu=\infty$) or the Euclidean norm ($\nu=2$). Define $U_\varepsilon(\thetao)$
and $\overline S_r(\varepsilon)$ as above, but with $\|\cdot\|_\nu$ in place of
$\|\cdot\|_\infty$. There exist $\varepsilon_0>0$ and a measurable
$\dot\ell_r\in L^2(P_{\thetao})$ such that, for every
$0<\varepsilon<\varepsilon_0$,
\[
|\ell_{r,\thet_1}(\xv)-\ell_{r,\thet_2}(\xv)|
\le
\dot\ell_r(\xv)\|\thet_1-\thet_2\|_\nu
\]
for all
\[
\xv\in\overline S_r(2\varepsilon),
\qquad
\thet_1,\thet_2\in U_\varepsilon(\thetao).
\]
\end{lemma}

\paragraph{Connection with the general moving-support likelihood theorem.}
The preceding results are organized to verify the four conditions of
\citet[Proposition~2.1]{buchersegers2017maximum}, which gives
asymptotic linearity of a maximum likelihood estimator in a parametric
model with parameter-dependent support.  In the notation of the present
paper, these conditions are:

\begin{enumerate}[label=(\alph*)]
	\item the model is differentiable in quadratic mean at $\thetao$ and
	$I_r(\thetao)$ is nonsingular;
	
	\item the probability of leaving the common support satisfies
	\[
	P_{\thetao}
	\{S_{r,\thetao}\setminus\overline S_r(\varepsilon)\}
	=o(\varepsilon^2);
	\]
	
	\item on the common support, the log-likelihood satisfies a local
	Lipschitz bound with an $L^2(P_{\thetao})$ envelope;
	
	\item the maximum likelihood estimator satisfies
	\[
	\sqrt n(\widehat\thet_n-\thetao)=O_p(1).
	\]
\end{enumerate}

For the fixed-$r$ $r$GEV family, condition~(a) is established by
Theorem~\ref{thm:dqm} and Proposition~\ref{prop:info-pd},
condition~(b) by Lemma~\ref{lem:support-loss},
condition~(c) by Lemma~\ref{lem:loglik-lipschitz}, and
condition~(d) by Proposition~\ref{prop:rootn}.

\begin{theorem}[Asymptotic linearity and normality]
\label{thm:asymp-linear}
Under the assumptions of Proposition~\ref{prop:rootn},
\begin{equation}
\sqrt n(\widehat\thet_n-\thetao)
=
I_r(\thetao)^{-1}
\frac1{\sqrt n}\sum_{i=1}^n
\dot{\boldsymbol\ell}_{r,\thetao}(\Xv_i)
+o_p(1).
\label{eq:asymp-linear}
\end{equation}
Consequently,
\[
\sqrt n(\widehat\thet_n-\thetao)
\rightsquigarrow
N_3\{0,I_r(\thetao)^{-1}\}.
\]
\end{theorem}

\begin{proof}[Proof of Theorem~\ref{thm:asymp-linear}]
	We apply \citet[Proposition~2.1]{buchersegers2017maximum} to the $r$GEV
	model, with sample space $\mathbb R^r$, Lebesgue measure as dominating
	measure and parameter set $K$. Since $\thetao$ is an interior point of $K$,
	sufficiently small balls around $\thetao$ are the same as in the
	unrestricted model. Conditions (a)--(d) above hold as indicated there. The
	norm in that proposition is not specified; we use the Euclidean norm. For
	condition~(b), a Euclidean $\varepsilon$-ball is contained in the
	maximum-norm $\varepsilon$-ball, so its common support is larger and
	Lemma~\ref{lem:support-loss} applies, and condition~(c) is
	Lemma~\ref{lem:loglik-lipschitz} with $\nu=2$. Hence \eqref{eq:asymp-linear}
	holds.

	The score $\dot{\boldsymbol\ell}_{r,\thetao}$ has mean zero and covariance
	$I_r(\thetao)$ under differentiability in quadratic mean. Since the block
	vectors are i.i.d.\ and $r$ is fixed, the multivariate central limit
	theorem and Slutsky's theorem give
	$\sqrt n(\widehat\thet_n-\thetao)\rightsquigarrow N_3\{0,I_r(\thetao)^{-1}\}$.
\end{proof}

Combined with the local asymptotic normality of the model
\citep[Corollary~3.4]{ParkShinPark2026A}, Theorem~\ref{thm:asymp-linear}
shows that no other estimator does better in the usual local sense. Recall that an
estimator sequence $T_n$ of $\psi(\thetao)$ is \emph{regular} at $\thetao$ if, for
every $\bm h\in\R^3$, the law of
$\sqrt n\{T_n-\psi(\thetao+\bm h/\sqrt n)\}$ under $P^n_{\thetao+\bm h/\sqrt n}$
converges to a limit that does not depend on $\bm h$, and \emph{best regular} if
this limit is $N\{0,\dot\psi_{\thetao}I_r(\thetao)^{-1}\dot\psi_{\thetao}^T\}$, the
smallest limit allowed by the convolution theorem
\citep[Theorem~8.8]{vanderVaart1998}.

\begin{corollary}[Asymptotic efficiency]
\label{cor:efficient}
Under the assumptions of Proposition~\ref{prop:rootn}, let $\psi$ be an
$\R^m$-valued map defined on $K$ and differentiable at $\thetao$. Then
$\psi(\widehat\thet_n)$ is a best regular estimator of $\psi(\thetao)$ at
$\thetao$. In particular, $\widehat\thet_n$ is regular and, for every $\bm h$,
\[
\sqrt n\bigl(\widehat\thet_n-\thetao-\bm h/\sqrt n\bigr)
\rightsquigarrow N_3\{0,I_r(\thetao)^{-1}\}
\qquad\text{under }P^n_{\thetao+\bm h/\sqrt n}.
\]
\end{corollary}

\begin{proof}
By Theorem~\ref{thm:asymp-linear} and the delta method
\citep[Theorem~3.1]{vanderVaart1998},
\[
\sqrt n\{\psi(\widehat\thet_n)-\psi(\thetao)\}
=\frac1{\sqrt n}\sum_{i=1}^n\dot\psi_{\thetao}I_r(\thetao)^{-1}
\dot{\boldsymbol\ell}_{r,\thetao}(\Xv_i)+o_p(1).
\]
For a model that is differentiable in quadratic mean at $\thetao$ with nonsingular
information, Lemma~8.14 of \citet{vanderVaart1998} states that an estimator
sequence is best regular for $\psi(\thetao)$ if and only if it admits exactly this
expansion. Its assumptions hold by Theorem~\ref{thm:dqm} and
Proposition~\ref{prop:info-pd}.
\end{proof}

By Corollary~\ref{cor:lan}(ii), the covariance
$\dot\psi_{\thetao}I_r(\thetao)^{-1}\dot\psi_{\thetao}^T$ is also the local
asymptotic minimax lower bound for every estimator sequence. The most important
example in applications is the return level. For $0<p<1$ put
$y_p=-\log(1-p)$ and let
\[
z_p(\thet)=
\begin{cases}
\mu+\sigma(1-y_p^{\,k})/k,&k\ne0,\\
\mu-\sigma\log y_p,&k=0,
\end{cases}
\]
be the $(1-p)$-quantile of the block maximum, so that $F_{\thet}\{z_p(\thet)\}=1-p$;
with annual blocks, $z_{0.01}$ is the 100-year return level.
The map $\thet\mapsto z_p(\thet)$ is real-analytic, also across $k=0$, with
gradient
\begin{equation}
\nabla z_p(\thet)=
\begin{pmatrix}
1\\[1mm]
(1-y_p^{\,k})/k\\[1mm]
-\sigma(1-y_p^{\,k})/k^2-\sigma y_p^{\,k}\log y_p/k
\end{pmatrix}
\quad(k\ne0),
\qquad
\nabla z_p(\thet)=
\begin{pmatrix}
1\\-\log y_p\\-\tfrac12\sigma(\log y_p)^2
\end{pmatrix}
\quad(k=0).
\label{eq:return-level-gradient}
\end{equation}

\begin{corollary}[Return levels]
\label{cor:return-level}
Under the assumptions of Proposition~\ref{prop:rootn}, for every $0<p<1$,
\[
\sqrt n\{z_p(\widehat\thet_n)-z_p(\thetao)\}
\rightsquigarrow
N\bigl\{0,\ \nabla z_p(\thetao)^TI_r(\thetao)^{-1}\nabla z_p(\thetao)\bigr\},
\]
and $z_p(\widehat\thet_n)$ is a best regular estimator of $z_p(\thetao)$.
\end{corollary}

\begin{proof}
Apply Corollary~\ref{cor:efficient} with $\psi=z_p$, using
\eqref{eq:return-level-gradient}.
\end{proof}

The limiting standard deviations in Theorem~\ref{thm:asymp-linear} and
Corollary~\ref{cor:return-level} are the efficient asymptotic standard deviations
evaluated numerically in \citet[Section~4]{ParkShinPark2026A}, and the
corresponding asymptotic standard errors of the maximum likelihood estimator are
these values divided by $\sqrt n$. Hence the maximum likelihood estimator attains
the efficiency gains reported there: from $r=1$ to $r=5$ its asymptotic standard
error decreases by about $45\%$ for the shape parameter and by $20$--$24\%$ for
the 100-year return level ($p=0.01$ with annual blocks) in heavy-tailed cases
$k_0\le-0.2$, but only by about $2\%$ for the return level when $k_0=0.4$.

\begin{remark}[Information identity and observed information]
\label{rem:observed-info}
By Theorem~\ref{thm:info-identity}, the limiting covariance in
Theorem~\ref{thm:asymp-linear} can equivalently be written as
$\{-\Ep_{\thetao}\ddot{\boldsymbol\ell}_{r,\thetao}\}^{-1}$. In practice this
covariance is estimated by the inverse observed information
$\{-P_n\ddot{\boldsymbol\ell}_{r,\widehat\thet_n}\}^{-1}$. Its consistency
requires, in addition, a uniform law of large numbers for the Hessian over a
shrinking neighborhood of $\thetao$, in which the moving support must again be
controlled; this is part of the ongoing work mentioned in
Section~\ref{sec:introduction}.
\end{remark}

\section{Finite-sample behavior}
\label{sec:simulation}

To see how well the normal approximation of Theorem~\ref{thm:asymp-linear} and
Corollary~\ref{cor:return-level} describes the maximum likelihood estimator at
realistic sample sizes, we simulated $n\in\{50,200\}$ independent blocks from the
$r$GEV model with $\thetao=(0,1,k_0)$ and $r\in\{1,3,5,7\}$, with $2000$
replications for each setting. Since heavy-tailed distributions are the most
relevant ones for environmental and financial extremes, we put the emphasis on
$k_0\in\{-0.4,-0.3,-0.2\}$, and add $k_0\in\{0,0.2,0.4\}$ for comparison. In each
replication the log-likelihood was maximized over $\{|k|<0.95,\ \sigma>0\}$ by the
Nelder--Mead method, started from the Gumbel moment estimates of $(\mu,\sigma)$
combined with each of $k\in\{-0.2,0,0.2,0.4\}$; the best of the four solutions was
retained. No replication failed, and the constraint $|k|<0.95$ was never active.

For the shape parameter we report the bias of $\widehat k_n$; for the 100-year
return level we report the relative bias of $z_{0.01}(\widehat\thet_n)$ in
percent. For both we report the ratio of $\sqrt n$ times the empirical standard
deviation to the efficient asymptotic standard deviation
$\{\dot\psi_{\thetao}I_r(\thetao)^{-1}\dot\psi_{\thetao}^T\}^{1/2}$; a ratio near one
means that the asymptotic standard error is accurate. The asymptotic values for
$r\le5$ are those of \citet[Table~1]{ParkShinPark2026A}. For $r=7$ they are
$0.461$, $0.431$, $0.400$, $0.339$, $0.276$ and $0.211$ for $k$, and $28.39$,
$19.05$, $12.85$, $5.90$, $2.63$ and $0.93$ for $z_{0.01}$, at
$k_0=-0.4,-0.3,-0.2,0,0.2,0.4$. The true return levels are $13.24$, $9.92$,
$7.55$, $4.60$, $3.01$ and $2.10$.

\begin{table}[!htb]
\centering
\caption{Simulation results for the maximum likelihood estimator ($2000$
replications, $\mu_0=0$, $\sigma_0=1$): bias of $\widehat k_n$, relative bias (\%)
of $z_{0.01}(\widehat\thet_n)$, and ratio of $\sqrt n$ times the empirical
standard deviation to the efficient asymptotic standard deviation.}
\label{tab:sim}
\small
\begin{tabular}{@{}rr cccc c cccc@{}}
\toprule
&& \multicolumn{4}{c}{$n=50$} && \multicolumn{4}{c}{$n=200$}\\
\cmidrule{3-6}\cmidrule{8-11}
&& \multicolumn{2}{c}{shape $k$} & \multicolumn{2}{c}{$z_{0.01}$} &&
\multicolumn{2}{c}{shape $k$} & \multicolumn{2}{c}{$z_{0.01}$}\\
$k_0$ & $r$ & bias & ratio & rel.\ bias & ratio && bias & ratio & rel.\ bias & ratio\\
\midrule
$-$0.4 & 1 & $-$0.008 & 1.15 & 12.5 & 1.42 && $-$0.003 & 1.04 & 2.9 & 1.09 \\
 & 3 & $-$0.003 & 1.09 & 8.3 & 1.25 && 0.000 & 0.99 & 1.3 & 1.02 \\
 & 5 & 0.002 & 1.03 & 4.2 & 1.11 && $-$0.001 & 0.99 & 1.8 & 1.03 \\
 & 7 & 0.000 & 1.02 & 4.4 & 1.07 && 0.001 & 1.01 & 0.6 & 1.03 \\
\addlinespace
$-$0.3 & 1 & 0.001 & 1.19 & 6.8 & 1.26 && $-$0.001 & 1.07 & 1.5 & 1.09 \\
 & 3 & 0.000 & 1.04 & 4.3 & 1.11 && 0.001 & 1.02 & 0.5 & 1.03 \\
 & 5 & 0.001 & 1.04 & 3.5 & 1.10 && 0.001 & 0.99 & 0.6 & 1.01 \\
 & 7 & $-$0.001 & 1.04 & 3.9 & 1.12 && 0.000 & 1.00 & 0.6 & 1.00 \\
\addlinespace
$-$0.2 & 1 & $-$0.005 & 1.17 & 6.1 & 1.26 && $-$0.003 & 1.03 & 1.5 & 1.04 \\
 & 3 & 0.004 & 1.06 & 1.9 & 1.09 && 0.001 & 0.98 & 0.1 & 0.98 \\
 & 5 & 0.004 & 1.05 & 1.3 & 1.09 && 0.001 & 1.01 & 0.3 & 1.02 \\
 & 7 & 0.002 & 1.02 & 2.0 & 1.04 && 0.000 & 1.00 & 0.8 & 1.01 \\
\addlinespace
0.0 & 1 & 0.008 & 1.19 & 0.0 & 1.13 && 0.002 & 1.04 & $-$0.3 & 1.02 \\
 & 3 & 0.003 & 1.09 & 0.3 & 1.09 && 0.001 & 1.04 & $-$0.1 & 1.03 \\
 & 5 & 0.004 & 1.09 & $-$0.3 & 1.07 && 0.002 & 1.01 & $-$0.5 & 1.01 \\
 & 7 & 0.004 & 1.07 & $-$0.1 & 1.06 && 0.001 & 1.01 & $-$0.1 & 1.01 \\
\addlinespace
0.2 & 1 & 0.021 & 1.34 & $-$3.4 & 1.16 && 0.006 & 1.12 & $-$1.0 & 1.03 \\
 & 3 & 0.011 & 1.18 & $-$2.4 & 1.12 && 0.003 & 1.07 & $-$0.6 & 1.05 \\
 & 5 & 0.009 & 1.13 & $-$2.3 & 1.08 && 0.002 & 1.06 & $-$0.6 & 1.05 \\
 & 7 & 0.006 & 1.15 & $-$1.6 & 1.08 && 0.003 & 1.03 & $-$0.8 & 1.03 \\
\addlinespace
0.4 & 1 & 0.035 & 1.66 & $-$4.5 & 1.57 && 0.012 & 1.36 & $-$1.5 & 1.33 \\
 & 3 & 0.016 & 1.39 & $-$3.4 & 1.42 && 0.006 & 1.24 & $-$1.1 & 1.27 \\
 & 5 & 0.014 & 1.35 & $-$3.4 & 1.42 && 0.005 & 1.23 & $-$1.3 & 1.30 \\
 & 7 & 0.010 & 1.32 & $-$2.7 & 1.40 && 0.003 & 1.17 & $-$1.1 & 1.28 \\
\bottomrule
\end{tabular}
\end{table}

\emph{Heavy tails.} For $k_0\in\{-0.4,-0.3,-0.2\}$ the asymptotic approximation
is accurate at $n=200$ for every $r$: the standard deviation ratios lie between
$0.98$ and $1.09$, and the relative bias of the return level is at most $3\%$. At
$n=50$ the block-maximum estimator ($r=1$) of the return level is noticeably
biased upwards and more variable than the asymptotic theory predicts, with
relative biases of $6$--$13\%$ and ratios of $1.26$--$1.42$. Additional order
statistics remove most of this discrepancy: with $r=5$ or $7$ the relative bias
falls to $1.3$--$4.4\%$ and the ratio to $1.04$--$1.12$. At $k_0=-0.4$, for example, the
relative bias decreases from $12.5\%$ for $r=1$ to $4.4\%$ for $r=7$, and the ratio
from $1.42$ to $1.07$. Thus, in heavy-tailed cases, the lower order statistics
improve not only the asymptotic efficiency quantified in
Corollary~\ref{cor:return-level}, but also the accuracy of the normal
approximation for small numbers of blocks. The shape estimator behaves well
throughout, with ratios between $1.02$ and $1.04$ for $r=7$ already at $n=50$.

\emph{Light and short tails.} For $k_0\in\{0,0.2\}$ the ratios at $n=200$ lie
between $1.01$ and $1.12$, and at $n=50$ they again tend to decrease with $r$.
Close to the regularity threshold, at $k_0=0.4$, the empirical standard deviations
still exceed the asymptotic values by $17$--$36\%$ at $n=200$. This is consistent
with the theory: the score is square integrable for $k_0<1/2$, but its third
absolute moment contains $\Ep T_1^{-3k_0}$, which is infinite for $k_0\ge1/3$, so
that the normal approximation cannot be expected to be accurate at the usual
$n^{-1/2}$ rate near the threshold. In this range the asymptotic standard errors
should be used with caution unless $n$ is large.

\section{Discussion}
\label{sec:discussion}

\subsection{Relation to the ordinary GEV theory}

Our arguments rely throughout on techniques developed for the GEV model. The
moving-support theorem, the modified criterion, the common-support localization
and the observation-dependent admissible sets are those of
\citet{buchersegers2017maximum}, and the consistency proof follows the Wald-type
argument of \citet{dombry2015existence}. What is new here is their
implementation for the ordered $r$-dimensional likelihood, which is not a
direct consequence of the GEV analysis. The observation is
an ordered dependent vector, and both the likelihood and its support involve
several order statistics. The parameter-free arrival representation
\eqref{eq:arrival} reduces this structure to exponential spacings and shows that
only the leading maximum or the smallest retained order statistic can reach the
moving boundary. The root-$n$ argument requires different devices in different
shape regimes: for $0<k_0<1/2$ a sequential modified criterion built from the
marginal and conditional likelihood factors, and for $k_0\le0$ a joint criterion
whose quadratic drift is reduced to that of the first marginal by the
data-processing inequality. The sequential criterion is a proof device, not a new
estimator: it is evaluated at the maximum likelihood estimator, which need not
maximize it. The same strategy should apply to other fixed-dimensional
order-statistic likelihoods whose supports are described by finitely many smooth
endpoint inequalities and whose local scores have an $L^2$ envelope.

We have aimed to keep the proofs checkable by hand. Through the arrival
representation \eqref{eq:arrival}, every probabilistic bound reduces to moments
of Gamma variables and exponential spacings, and apart from standard
empirical-process results \citep{vanderVaart1998} the analytic steps use only
Taylor expansions of $T_{\thet}$ and elementary inequalities.
Appendix~\ref{app:consistency} proves consistency, Appendix~\ref{app:rate}
the root-$n$ rate, and Appendix~\ref{app:support} the support-loss and
Lipschitz conditions, so that each lemma can be matched to the condition of
\citet[Proposition~2.1]{buchersegers2017maximum} that it verifies.

\subsection{Scope and extensions}

The efficiency result (Corollary~\ref{cor:efficient}) also gives a benchmark for
other estimators of the $r$GEV parameters. Robust procedures such as that of
\citet{dupuis1997extreme} trade some efficiency at the model for protection
against outliers, and $I_r(\thetao)^{-1}$ quantifies this price.

Between the two thresholds, for $1/2\le k_0<1$, Theorem~\ref{thm:consistency}
still gives consistency, but the model is not differentiable in quadratic mean
\citep{ParkShinPark2026A}. In the leading coordinate the $r$GEV density
has the endpoint form $c(x-\vartheta)^{\alpha-1}$ with $\alpha=1/k_0$ studied by
\citet[Theorem~3 and Section~7]{Smith1985}, who showed that the classical
$\sqrt n$ normal theory applies when $\alpha>2$ but not when $\alpha\le2$. We
expect the same non-regular behavior for every fixed $r$, since only $X^{(1)}$ approaches the endpoint, but do not pursue this
here.

Allowing $r=r_n\to\infty$ would create a growing-dimensional observation vector and a
sequence of information matrices $I_{r_n}(\thetao)$.  The appropriate normalization could
then depend on the rate at which information grows with $r_n$, and uniform versions of the
support and empirical-process arguments would be required.  If the $r$GEV model is
viewed as an approximation arising from blocks of an underlying sample, the rate of growth
of $r_n$ would additionally have to be related to the underlying block size, as in the
block-maxima analysis of \citet{dombry2019maximum}.

A second extension is data-adaptive selection of $r$.  If
\[
P(\widehat r_n=r_0)\longrightarrow1
\]
for a fixed finite $r_0$, the present results supply the fixed-model component needed
for an oracle-type post-selection argument.  When the limiting selection probabilities
do not collapse to a single value, however, the distribution of
$\widehat\thet_{n,\widehat r_n}$ may depend on the selection event and need not reduce to
a single normal law.  This question is particularly relevant to procedures that
choose how many largest order statistics to retain from the data
\citep{bader2017automated,ShinParkPark2026}. For the procedure of
\citet{ShinParkPark2026}, the present results supply the consistency and
asymptotic normality assumed there for every fixed candidate $r$ whenever
$k_0<1/2$.

Finally, the exact-model assumptions made here separate the behavior of the estimator
from dependence and finite-block approximation.  \citet{smith1986extreme} already
emphasized that $r$ should remain small when the asymptotic largest-order-statistic
approximation is used in practice, and \citet{BucherHaufs2026} show in a top-two
time-series setting that naive independence-based likelihood procedures can fail
under serial dependence and develop bias-corrected alternatives.  Extending the
present theory to dependent blocks remains open.

\section{Conclusion}

For the exact three-parameter $r$GEV model with a fixed number $r$ of order
statistics, the maximum likelihood estimator is strongly consistent on compact
parameter sets contained in $k<1$. Whenever $k_0<1/2$, the range in which the
model is regular, it is root-$n$ consistent, asymptotically linear with influence
function $I_r(\thetao)^{-1}\dot{\boldsymbol\ell}_{r,\thetao}$, asymptotically normal
and asymptotically efficient. The regularity threshold is the same for every
fixed $r$ as for the block-maximum GEV model, because only the leading maximum
or the smallest retained order statistic can leave the support of a shrinking
parameter neighborhood. These results justify the large-sample normal
approximation that is routinely used with $r$-largest models, and they are the
basis for likelihood-based inference with a fixed number of upper order
statistics.

\begin{small}
\renewcommand{\baselinestretch}{1.0}\selectfont
\setlength{\parskip}{3pt}

\subsection*{Statements and Declarations}

\noindent\textbf{ORCID.}
Yire Shin: 0000-0003-1297-5430; Jihong Park: 0009-0003-3191-9968;
Jeong-Soo Park: 0000-0002-8460-4869.

\noindent\textbf{Funding.}
Jihong Park's research was supported by the BK21 FOUR (Fostering Outstanding
Universities for Research, No.~5120200913674), funded by the Ministry of
Education (MOE, Korea) and the National Research Foundation of Korea (NRF).
Yire Shin's research was supported by the Basic Science Research Program
through the National Research Foundation of Korea (NRF), funded by the Ministry
of Education (RS-2025-25436608).

\noindent\textbf{Conflict of interest.}
The authors declare that they have no conflicts of interest.

\noindent\textbf{Use of generative artificial intelligence.}
Generative artificial intelligence tools, namely ChatGPT (OpenAI) and
Claude (Anthropic), were used during this work. They assisted with
exploring proof strategies, drafting and checking mathematical derivations,
and editing the manuscript for clarity. The research questions and the key ideas, including the sequential decomposition of the Fisher information, originated with the
authors. All statements, proofs, references and conclusions were checked by the authors, who take full responsibility for the content of this article.

\noindent\textbf{Author contributions.}
All authors contributed to the conception of the study, development of the
methodology, and derivation and verification of the theoretical results.
The first draft of the manuscript was written by Shin and J.-S. Park, and
the other authors commented on and revised previous versions of the manuscript.
All authors read and approved the final manuscript.
\par
\end{small}

\appendix
\section*{Appendix}

\paragraph{Sign convention and sources.}
Throughout the appendix, $k$ denotes the shape parameter in
\eqref{eq:shape-sign-conversion}. Results of \citet{buchersegers2017maximum} and
\citet{dombry2015existence}, stated there in the shape parameter $\gamma$, are used
after the substitution $\gamma=-k$; in particular $\gamma>-1/2$ if and only if
$k<1/2$. The map $(\mu,\sigma,k)\mapsto(-k,\mu,\sigma)$ is a linear bijection that
preserves both the maximum and the Euclidean norms. Lipschitz bounds and
second-order expansions therefore transfer unchanged, and the Fisher information
transforms as $I^{(k)}=J^TI^{(\gamma)}J$ with $J$ the corresponding signed
permutation matrix. The modified criterion, the common-support localization and
the observation-dependent admissible sets are taken from
\citet[Section~2 and Appendices~D--E]{buchersegers2017maximum}. Their
implementation for the ordered $r$-dimensional likelihood, including the
sequential criterion of Appendix~\ref{app:rate}, is new.

\section{Proof of strong consistency}
\label{app:consistency}

\begin{lemma}[Integrability of the true log-likelihood]
\label{lem:true-loglik-integrable}
For fixed $r$ and every interior true parameter $\thetao$,
\[
\Ep_{\thetao}|\ell_{r,\thetao}(\Xv)|<\infty.
\]
\end{lemma}

\begin{proof}
By \eqref{eq:loglik-T},
\[
\ell_{r,\thetao}
=-r\log\sigma_0-T_r+(1-k_0)\sum_{j=1}^r\log T_j.
\]
By \eqref{eq:arrival}, $T_j\sim\operatorname{Gamma}(j,1)$, so $T_r$ has finite
mean and
\[
\Ep|\log T_j|=\frac1{\Gamma(j)}
\int_0^\infty |\log t|\,t^{j-1}e^{-t}\,dt<\infty .
\]
The claim follows.
\end{proof}

\begin{proof}[Proof of Theorem~\ref{thm:consistency}]
For $\thet\in K\subset\Theta_1$,
\[
\ell_{r,\thet}
=-r\log\sigma-T_r+(1-k)\sum_{j=1}^r\log T_j
\le
-r\log\sigma-T_r+r(1-k)\log T_r.
\]
Because $K$ is compact, $\sigma$ is bounded away from zero and $1-k$ ranges in a
compact subset of $(0,\infty)$.  Since $-t+a\log t$ is bounded above uniformly for
$t>0$ and $a$ in a compact positive interval,
\begin{equation}
\sup_{\thet\in K,\,\xv}\ell_{r,\thet}(\xv)<\infty.
\label{eq:uniform-upper-loglik}
\end{equation}

In the extended-real sense,
\[
P_{\thetao}\ell_{r,\thet}-P_{\thetao}\ell_{r,\thetao}
=-D_{\mathrm{KL}}(P_{r,\thetao}\Vert P_{r,\thet})\le0.
\]
Equality implies equality of the joint distributions and therefore equality of their first
marginals.  The first marginal is the ordinary GEV distribution, which is identifiable;
hence equality occurs only at $\thet=\thetao$.

For a parameter set $B\subset K$, put $\ell_B(\xv)=\sup_{\thet\in B}\ell_{r,\thet}(\xv)$. If $B$ is relatively open in $K$, then $\ell_B$ is measurable. Indeed,
$\{\thet\in B:\xv\in S_{r,\thet}\}$ is relatively open in $B$ and
$\thet\mapsto\ell_{r,\thet}(\xv)$ is continuous on it, so $\ell_B(\xv)$ equals the
supremum over a fixed countable dense subset of $B$. For each fixed $\xv$, the extended-real map $\thet\mapsto\ell_{r,\thet}(\xv)$,
with $\ell_{r,\thet}(\xv)=-\infty$ outside $S_{r,\thet}$, is upper
semicontinuous; compare \citet{dombry2015existence} for the GEV
likelihood. Inside the support this is ordinary continuity. If $\xv$ lies
strictly outside $S_{r,\thet}$, it also lies outside $S_{r,\thet'}$ for all
$\thet'$ near $\thet$, and both values are $-\infty$. If $\xv$ lies on the
boundary of $S_{r,\thet}$, the log-density tends to $-\infty$ as that boundary
is approached: at a finite upper endpoint ($k>0$), $T_1\downarrow0$ and
$(1-k)\log T_1\to-\infty$ because $k<1$; at a finite lower endpoint ($k<0$),
$T_r\to\infty$ and the term $-T_r$ dominates. Writing $B(\thet,\eta)$ for the
open ball of radius $\eta$ about $\thet$, upper semicontinuity gives
\[
\ell_{B(\thet,\eta)}(\xv)\downarrow\ell_{r,\thet}(\xv)
\qquad(\eta\downarrow0).
\]
Let $C$ be the uniform upper bound in
\eqref{eq:uniform-upper-loglik}. As $\eta$ decreases, the functions
$C-\ell_{B(\thet,\eta)}$ are nonnegative and increase to $C-\ell_{r,\thet}$.
The monotone convergence theorem \citep[Chapter~2]{Folland1999}, which
allows the limit to be $+\infty$, therefore gives $P_{\thetao}\ell_{B(\thet,\eta)}\longrightarrow P_{\thetao}\ell_{r,\thet}$. Hence, for every $\thet\ne\thetao$, a sufficiently small neighborhood $B_{\thet}$ satisfies $P_{\thetao}\ell_{B_{\thet}}<P_{\thetao}\ell_{r,\thetao}$.

Fix $\delta>0$.  The compact set $K_\delta=\{\thet\in K:\|\thet-\thetao\|\ge\delta\}$ admits a finite cover $B_1,\ldots,B_m$ of such open neighborhoods; this is
the Wald-type argument of \citet[Theorem~5.14]{vanderVaart1998} and
\citet[proofs of Theorem~2 and Proposition~2]{dombry2015existence}; see also
\citet[proof of Proposition~3.1]{buchersegers2017maximum}.
For each $j$ and each integer $M>0$, the truncation
$\max(\ell_{B_j},-M)$ is bounded, so the strong law gives
$\limsup_n P_n\ell_{B_j}\le P_{\thetao}\max(\ell_{B_j},-M)$ almost surely.
Letting $M\to\infty$ yields
\[
\limsup_nP_n\ell_{B_j}\le P_{\thetao}\ell_{B_j}
\qquad\text{a.s.}
\]
Therefore
\[
\limsup_n\sup_{\thet\in K_\delta}P_n\ell_{r,\thet}
< P_{\thetao}\ell_{r,\thetao}
\qquad\text{a.s.}
\]
whereas Lemma~\ref{lem:true-loglik-integrable} yields $P_n\ell_{r,\thetao}\to P_{\thetao}\ell_{r,\thetao} \quad\text{a.s.}$ Let $\Omega_0$ be the intersection of these events over $\delta=1/q$,
$q=1,2,\ldots$; it has probability one and does not depend on the choice of
maximizers. On $\Omega_0$, for every $\delta=1/q$, any maximizer over $K$ lies
outside $K_\delta$ for all sufficiently large $n$, because its criterion value is
at least $P_n\ell_{r,\thetao}$. Hence every sequence of maximizers converges to
$\thetao$ on $\Omega_0$. Finally, maximizers exist for every sample, because
$\thet\mapsto P_n\ell_{r,\thet}$, a finite sum of upper semicontinuous functions
with values in $[-\infty,\infty)$, attains its supremum on the compact set $K$.
\end{proof}

\section{Root-$n$ rate}
\label{app:rate}

We use the modified-criterion method of
\citet[Section~2 and Appendix~D]{buchersegers2017maximum}. For $k_0\le0$ the
criterion is applied directly to the joint density $p_{r,\thet}$
(Lemmas~\ref{lem:joint-mod-negative} and \ref{lem:joint-mod-gumbel}). For
$0<k_0<1/2$ it is applied separately to the marginal and conditional factors
of \eqref{eq:sequential-factorization} (Lemmas~\ref{lem:seq-criterion}--\ref{lem:seq-drift}).
The reason is the moving upper endpoint. By \eqref{eq:modified-density-power}
below, the Lipschitz envelope of a modified criterion is built from
$\sup_{\thet}\|\nabla_{\thet}p^a_{\thet}\|/p^a_{\thetao}$ for some power
$a\in(0,1]$. Near an upper endpoint, where $v=T_{\thet}(x)\downarrow0$, the density
behaves like $v^{1-k}$, so $p^a_{\thet}\approx v^{a(1-k)}$, and differentiation in
$\thet$ adds a factor of order $v^{-k}$ by \eqref{eq:D-uniform}. The gradient is
therefore of order $v^{a(1-k)-k}$; it tends to zero, which makes the zero
extension continuously differentiable, when $a>k/(1-k)$. In the lower tail of the
last coordinate, on the other hand, the numerator contains
$\exp\{-aT_{\thet}(x_r)\}$ and the denominator contributes
$\exp\{aT_{\thetao}(x_r)\}$. Since $T_{\thet}(x_r)/T_{\thetao}(x_r)$ does not tend to
one in that tail unless $k=k_0$, the two exponentials cannot be compared, and we
bound the numerator by a polynomial; square integrability against
$p_{\thetao}\approx e^{-T_{\thetao}(x_r)}$ then requires $a<1/2$. (With an
exponential lower bound for $T_{\thet}$, as in
\citet[Lemma~D.3, Case~II.2]{buchersegers2017maximum}, $a=1/2$ can also be used,
but not $a>1/2$.) For the GEV density,
\citet[Lemma~D.3]{buchersegers2017maximum} meet the two requirements by using
different powers on different intervals of the real line. For the joint density
does not work in the same way, because a single observation vector can have
$x_1$ close to the moving endpoint and $x_r$ far in the lower tail. With one
power, the envelope argument just described therefore works only when
$k/(1-k)<1/2$, that is, when $k<1/3$. When $0<k_0<1/2$ we
therefore apply the criterion to the one-dimensional conditional factors, where a
larger power can be used for observations near the endpoint and a power below
$1/2$ in the tail, and where the preceding arrival time supplies extra
integrability near the endpoint. The sequential criterion and the ordinary
criterion are evaluated at the same likelihood maximizer.

All Lipschitz bounds below rest on the density-power inequality obtained by the
argument in the proof of \citet[Lemma~D.3]{buchersegers2017maximum}, where it is
used for $a\in[1/2,1)$; the same argument gives it for all $a\in(0,1]$: for $p_0>0$,
$p_1,p_2\ge0$ and $a\in(0,1]$,
\begin{equation}
\left|2\log\frac{p_1+p_0}{2p_0}-2\log\frac{p_2+p_0}{2p_0}\right|
\le\frac2a\,\frac{|p_1^a-p_2^a|}{p_0^a}.
\label{eq:modified-density-power}
\end{equation}

Define the ordinary modified criterion
\[
m_{1,\thet}(x)
=2\log\left\{\frac{f_{\thet}(x)+f_{\thetao}(x)}{2f_{\thetao}(x)}\right\}.
\]
For the sequential construction, suppose $0<k_0<1/2$, fix
$0<k_-<k_0<k_+<1/2$, and let $V$ be an open convex neighborhood of $\thetao$
whose closure has $k\in[k_-,k_+]$ and $\sigma$ bounded away from zero and
infinity; $V$ is shrunk finitely many times in the proof of
Lemma~\ref{lem:cond-mod-positive}. For $\thet\in V$ we use the GEV distribution
function $F_{\thet}$ and density $f_{\thet}$ on the whole real line, with
$F_{\thet}=1$ and $f_{\thet}=0$ above the upper endpoint $\omega(\thet)$; then
$F_{\thet}(x)>0$ for every finite $x$.
For $s\ge2$, define
\[
q_{s,\thet}(y\mid x)
=\frac{f_{\thet}(y)}{F_{\thet}(x)}\ind(y<x)
\]
and
\[
m_{s,\thet}(x,y)
=2\log\left\{
\frac{q_{s,\thet}(y\mid x)+q_{s,\thetao}(y\mid x)}
{2q_{s,\thetao}(y\mid x)}
\right\}.
\]
Finally set
\begin{equation}
M_{r,\thet}(\xv)
=m_{1,\thet}(x_1)+\sum_{s=2}^r m_{s,\thet}(x_{s-1},x_s).
\label{eq:sequential-modified}
\end{equation}

\begin{lemma}[Sequential modified-criterion inequality]
\label{lem:seq-criterion}
If $\widehat\thet_n$ maximizes the $r$GEV likelihood over a set containing
$\thetao$ and $\widehat\thet_n\in V$, then $P_nM_{r,\widehat\thet_n}\ge P_nM_{r,\thetao}=0$. \end{lemma}

\begin{proof}
For $u\ge0$ and $w>0$, $2\log\frac{u+w}{2w}\ge\log\frac uw$ by the arithmetic--geometric mean inequality.  Apply this to every factor in
\eqref{eq:sequential-factorization} and sum.  Then $M_{r,\thet}(\xv)\ge \log\frac{p_{r,\thet}(\xv)}{p_{r,\thetao}(\xv)}$. Since $\widehat\thet_n$ maximizes the likelihood,
$P_nM_{r,\widehat\thet_n}\ge P_n\log(p_{r,\widehat\thet_n}/p_{r,\thetao})\ge0$;
note that $\widehat\thet_n$ need not maximize $M_{r,\thet}$ itself.
\end{proof}

\begin{lemma}[Conditional modified criterion for $0<k_0<1/2$]
\label{lem:cond-mod-positive}
Let $s\in\{2,\ldots,r\}$ and $0<k_0<1/2$. After shrinking $V$ if necessary,
there exists $\dot m_s\in L^2(P_{\thetao})$ such that, $P_{\thetao}$-almost surely,
\[
|m_{s,\thet_1}(X^{(s-1)},X^{(s)})-m_{s,\thet_2}(X^{(s-1)},X^{(s)})|
\le \dot m_s(X^{(s-1)},X^{(s)})\|\thet_1-\thet_2\|
\qquad\text{for all }\thet_1,\thet_2\in V.
\]
\end{lemma}

\begin{proof}
	We extend the argument of \citet[Lemma~D.3]{buchersegers2017maximum} to a
conditional factor, with a larger power $a_E$ near the finite endpoint and a
smaller power $a_T$ in the opposite tail. Fix $s\in\{2,\ldots,r\}$ and $0<k_0<1/2$. Put $t=T_{\thetao}(x),\quad u=T_{\thetao}(y),\quad 0<t<u$. 	Under $P_{\thetao}$, their joint density is
	\[
	g_s(t,u)=\frac{t^{s-2}e^{-u}}{\Gamma(s-1)}\mathbf1\{0<t<u\}.
	\]
	Let $k_\pm$ and $V$ be as above, and fix $\eta\in(0,1)$. For $k>0$, write $\omega(\thet)=\mu+\sigma/k$ for the upper endpoint and extend $T_{\thet}$ by
	\[
	\bar T_{\thet}(z)=
\begin{cases}
T_{\thet}(z),&z<\omega(\thet),\\
0,&z\ge \omega(\thet),
\end{cases}
	\qquad
	\bar\lambda_{\thet}(z)=\sigma^{-1}\bar T_{\thet}(z)^{1-k}\mathbf1\{z<\omega(\thet)\}.
	\]
	Since $k>0$, every finite $x$ has $F_{\thet}(x)>0$, with $F_{\thet}(x)=1$
above $\omega(\thet)$. Hence $q_{s,\thet}(\cdot\mid x)$ is a probability
density for every finite $x$, including conditioning values outside the
candidate support, and for $y<x$
	\[
	q_{s,\thet}(y\mid x)=\bar\lambda_{\thet}(y)
	\exp\{-\bar T_{\thet}(y)+\bar T_{\thet}(x)\}.
	\]
	On candidate support put $v=T_{\thet}(y)$ and $b=\bar T_{\thet}(x)$; monotonicity gives $0\le b\le v$. For $0<a<1$, let $Q_{a,\thet}=q_{s,\thet}^a$. Then
	\begin{equation}
	Q_{a,\thet}=\sigma^{-a}v^{a(1-k)}e^{-a(v-b)}
	\le C v^{a(1-k)}.
	\label{eq:cond-density-power}
	\end{equation}
	Writing $D_{\thet}(z)=\nabla_{\thet}\log T_{\thet}(z)$, differentiation of
	$T_{\thet}(z)^k=1-k(z-\mu)/\sigma$ shows, uniformly on $V$, that
	\begin{equation}
	\|D_{\thet}(z)\|\le C(1+v^{-k})(1+|\log v|),\quad
	\|\nabla_{\thet} T_{\thet}(z)\|
	\le Cv(1+v^{-k})(1+|\log v|),\qquad v=T_{\thet}(z)>0.
	\label{eq:cond-transform-derivatives}
	\end{equation}
	In particular, $\bar T_{\thet}(z)$ and its first parameter derivatives vanish continuously at the moving upper endpoint, because $k_+<1$. On the interior of candidate support,
	\begin{equation}
	\begin{aligned}
	\nabla_{\thet} Q_{a,\thet}
	=aQ_{a,\thet}\bigl[&-\sigma^{-1}e_\sigma+(1-k)D_{\thet}(y)
	-e_k\log v\\
	&-vD_{\thet}(y)+\bar T_{\thet}(x)D_{\thet}(x)\bigr],
	\end{aligned}
	\label{eq:cond-power-gradient}
	\end{equation}
	where the last term is $\nabla_{\thet}\bar T_{\thet}(x)$, which is zero for
$x\ge\omega(\thet)$.
	
	For $0<u\le\eta$, the representation
	\[
	v=h_{\thet}(u)=\{A_{\thet}+B_{\thet} u^{k_0}\}^{1/k}
	\]
	(on candidate support), with
	\[
	A_{\thet}=1-\frac{k}{\sigma}\left(\mu_0-\mu+\frac{\sigma_0}{k_0}\right),
	\qquad B_{\thet}=\frac{k\sigma_0}{\sigma k_0},
	\]
	is obtained by substituting $y=T_{\thetao}^{-1}(u)=\mu_0+(\sigma_0/k_0)(1-u^{k_0})$
into $w_{\thet}(y)=1-k(y-\mu)/\sigma$. Since $A_{\thet}$ and $B_{\thet}$ are
bounded on $V$ and $u\le\eta$, it shows that $v=T_{\thet}\circ T_{\thetao}^{-1}(u)$
is bounded above by a constant $C_\eta$, uniformly on $V$.
The power $a_E$ is chosen to satisfy the endpoint requirement $a>k/(1-k)$
explained at the start of this appendix; since $k_+<1/2$, the interval below
is nonempty, and we may choose
	\[
	\frac{k_+}{1-k_+}<a_E<1,\qquad
	\rho=a_E(1-k_+)-k_+>0.
	\]
	By \eqref{eq:cond-transform-derivatives}, the conditioning-value term
$\nabla_{\thet}\bar T_{\thet}(x)$ is bounded by $Cb(1+b^{-k})(1+|\log b|)$, which is
uniformly bounded for $0\le b\le v\le C_\eta$ because $k\le k_+<1$.
By \eqref{eq:cond-density-power}--\eqref{eq:cond-power-gradient},
$0\le b\le v$, and the boundedness of $v$,
	\[
	\|\nabla_{\thet} Q_{a_E,\thet}(x,y)\|
	\le Cv^\rho(1+|\log v|^2)\le C_E\quad(v\le1),
	\]
	with a constant bound for $1\le v\le C_\eta$. These bounds tend to zero as
$v\downarrow0$, and $\nabla_{\thet}\bar T_{\thet}(x)\to0$ as $b\downarrow0$.
Lemma~\ref{lem:zero-extension}, applied to the two factors of
$Q_{a_E,\thet}$ (with $m=1$ and $x_1=y$, $c=0$, respectively $x_1=x$,
$c=1$ for $e^{a_E\bar T_{\thet}(x)}$), shows that
$\thet\mapsto Q_{a_E,\thet}(x,y)$ is continuously differentiable on $V$,
even along parameter paths that cross either moving endpoint. Thus
	\begin{equation}
	\sup_{\thet\in V}\|\nabla_{\thet} Q_{a_E,\thet}(x,y)\|\le C_E
	\qquad(0<u\le\eta).
	\label{eq:cond-endpoint-bound}
	\end{equation}
	For $u>\eta$, shrink $V$ so that $B_{\thet}\ge B_->0$ and
	$|A_{\thet}|\le\tfrac12 B_-\eta^{k_0}$. Then $\tfrac12 B_-u^{k_0}\le A_{\thet}+B_{\thet} u^{k_0} \le C(1+u^{k_0})$. 	Hence $y$ lies inside every candidate support for $\thet\in V$, and $v$ and
$v^{-1}$ are bounded by fixed powers of $1+u$ uniformly over $V$. For fixed $0<a_T<1/2$, \eqref{eq:cond-density-power}--\eqref{eq:cond-power-gradient}, with $0\le b\le v$, therefore yield constants $C,M<\infty$ such that
	\begin{equation}
	\sup_{\thet\in V}\|\nabla_{\thet} Q_{a_T,\thet}(x,y)\|
	\le C(1+u)^M,\qquad u>\eta.
	\label{eq:cond-tail-bound}
	\end{equation}
	Only $x$ can cross the upper endpoint in this region, and
Lemma~\ref{lem:zero-extension} applies as before; the term
$\nabla_{\thet}\bar T_{\thet}(x)$ is bounded by a constant for $b\le1$ and by
$C(1+v)^2$ for $1<b\le v$.
	
	Inequality \eqref{eq:modified-density-power}, applied with $p_i=q_{s,\thet_i}(y\mid x)$,
	together with \eqref{eq:zero-extension-mvt}, gives the asserted Lipschitz
	inequality, with envelope bounded, up to constants, by
	\[
	\dot m_s(t,u)=C\begin{cases}
	u^{-a_E(1-k_0)},&0<u\le\eta,\\
	(1+u)^M u^{-a_T(1-k_0)}e^{a_T(u-t)},&u>\eta.
	\end{cases}
	\]
	Indeed, $q_{s,\thetao}(y\mid x)=\sigma_0^{-1}u^{1-k_0}e^{-(u-t)}$. In the endpoint region,
	\[
	\int_0^\eta\int_0^u t^{s-2}u^{-2a_E(1-k_0)}\,dt\,du
	=\frac1{s-1}\int_0^\eta u^{s-1-2a_E(1-k_0)}\,du<\infty,
	\]
	because $a_E(1-k_0)<1$ gives $s-1-2a_E(1-k_0)>s-3\ge-1$, which is the
condition for integrability of a power at zero. Here the factor $u^{s-1}$,
obtained by integrating out the preceding arrival time $t\in(0,u)$, is the
extra integrability supplied by a conditional factor. For the marginal
factor the same computation gives the exponent $-2a_E(1-k_0)$, which would
require the additional restriction $a_E<1/\{2(1-k_0)\}$; the marginal factor
is instead covered by \citet[Lemma~D.3]{buchersegers2017maximum}. In the tail region, $a_T<1/2$ and $e^{a_T(u-t)}\le e^{a_Tu}$ give
	\[
	\int_\eta^\infty(1+u)^{2M}u^{-2a_T(1-k_0)}
	e^{-(1-2a_T)u}\left(\int_0^u t^{s-2}\,dt\right)du<\infty.
	\]
	Thus $\dot m_s\in L^2(P_{\thetao})$, as required.
\end{proof}

\begin{lemma}[Quadratic drift for the sequential criterion]
\label{lem:seq-drift}
If $0<k_0<1/2$, there exist $c,\delta>0$ such that
\[
P_{\thetao}\{M_{r,\thet}-M_{r,\thetao}\}
\le-c\|\thet-\thetao\|^2
\]
whenever $\|\thet-\thetao\|<\delta$.
\end{lemma}

\begin{proof}
For every $s\ge2$ and fixed $x$, $q_{s,\thet}(\cdot\mid x)$ is a
probability density in the positive-shape neighborhood used here.
Its average with $q_{s,\thetao}$ is therefore a probability density too.
Nonnegativity of relative entropy gives
\[
P_{\thetao}\{m_{s,\thet}\mid X^{(s-1)}=x\}
=-2D_{\mathrm{KL}}\left(q_{s,\thetao}(\cdot\mid x)\,
\bigg\|\,\frac{q_{s,\thet}(\cdot\mid x)+q_{s,\thetao}(\cdot\mid x)}2\right)
\le0.
\]
Hence $P_{\thetao}M_{r,\thet}\le P_{\thetao}m_{1,\thet}$. For the first component, \citet[Lemma~D.4]{buchersegers2017maximum} gives,
after the change of parameters described at the start of the appendices,
\[
P_{\thetao}m_{1,\thetao+\bm h}
=-\frac14\bm h^TI_1(\thetao)\bm h+o(\|\bm h\|^2).
\]
Since $I_1(\thetao)>0$ by Proposition~\ref{prop:info-pd}, there are $c,\delta>0$
with $P_{\thetao}m_{1,\thet}\le-c\|\thet-\thetao\|^2$ for
$\|\thet-\thetao\|<\delta$, which proves the claim.
\end{proof}

For $k_0\le0$ it is convenient to use the joint modified criterion
\[
m^J_{r,\thet}(\xv)
=2\log\left\{
\frac{p_{r,\thet}(\xv)+p_{r,\thetao}(\xv)}{2p_{r,\thetao}(\xv)}
\right\}.
\]

\begin{lemma}[Joint modified criterion for $k_0<0$]
\label{lem:joint-mod-negative}
If $k_0<0$, there exist a neighborhood $V$ of $\thetao$ and
$\dot m_r^J\in L^2(P_{\thetao})$ such that
\[
|m^J_{r,\thet_1}-m^J_{r,\thet_2}|
\le \dot m_r^J\|\thet_1-\thet_2\|,
\qquad \thet_1,\thet_2\in V.
\]
\end{lemma}

\begin{proof}
	Choose an open convex neighborhood $V$ of $\thetao$ whose closure has
	$k$ negative and bounded away from zero, and $\sigma$ bounded above and
	away from zero. Put $s_{\thet}=p_{r,\thet}^{1/2}$, extended by zero outside
	its support. By Lemma~\ref{lem:sqrt-smooth} (here only a lower endpoint can move,
	since $k<0$ on $V$), $\thet\mapsto s_{\thet}(\xv)$ is
	continuously differentiable on $V$ for every $\xv$, with
	$\nabla s_{\thet}=\tfrac12s_{\thet}\dot{\boldsymbol\ell}_{r,\thet}$ on the
	support and $\nabla s_{\thet}=0$ off it. It remains to dominate the gradient.

		Use $t_j=T_{\thetao}(x_j)$ and
	\[
	x_j=\mu_0+\frac{\sigma_0}{k_0}(1-t_j^{k_0}),\qquad
	|dx_j|=\sigma_0t_j^{k_0-1}\,dt_j.
	\]
	On candidate support write
	\[
	h_{\thet}(t)=\{A_{\thet}+B_{\thet} t^{k_0}\}^{1/k},\qquad
	\rho(\thet)=k_0/k>0,
	\]
	with $A_{\thet},B_{\thet}$ as in Lemma~\ref{lem:cond-mod-positive}. The coefficients $A_{\thet}$
	are bounded and $B_{\thet}$ are bounded above and away from zero.
	There is a fixed $\delta\in(0,1)$ such that every point with
	$0<t\le\delta$ is supported for every $\thet\in V$, and
	\[
	h_{\thet}(t)\asymp t^{\rho(\thet)},\qquad
	0<h_{\thet}(t)\le1\quad(0<t\le\delta),
	\]
	uniformly on $V$, where $\asymp$ means that both ratios are bounded by
constants independent of $t$ and $\thet$. Indeed, $t^{k_0}\to\infty$ as
$t\downarrow0$, so $A_{\thet}+B_{\thet}t^{k_0}$ lies between two positive
constant multiples of $B_{\thet}t^{k_0}$, uniformly on $V$; raising to the power $1/k$, with $k$ in a
compact negative interval, gives the comparison.
Put $\rho_* =\inf_{\thet\in V}\rho(\thet)>0$.
	Let $\lambda_{\thet}=\sigma^{-1}T_{\thet}^{1-k}$ and
	$\Phi_{\thet}=1+\|D_{\thet}\|^2+|\log T_{\thet}|^2$.
	Formula \eqref{eq:D}, with $k$ bounded away from zero, gives $\|D_{\thet}(x)\|\le C(1+v^{-k})(1+|\log v|),\quad v=T_{\thet}(x)$. 	Consequently, for $t\le\delta$,
	\begin{equation}
	\lambda_{\thet}(x(t))\Phi_{\thet}(x(t))t^{k_0-1}
	\le Ct^{\rho_*-1}(1+|\log t|^4),                         \label{eq:joint-negative-small-t}
	\end{equation}
	where we used the cancellation
	$\rho(\thet)(1-k)+k_0-1=\rho(\thet)-1$.
	For $t\ge\delta$ on candidate support, monotonicity gives
	$h_{\thet}(t)\ge h_{\thet}(\delta)\ge c_\delta>0$.
	Thus $\lambda_{\thet} \Phi_{\thet}$ is bounded by a fixed finite power
	of $1+h_{\thet}(t)$ in this region.
	
	By the Cauchy--Schwarz inequality, the squared norm of \eqref{eq:joint-score}
is at most a constant times the sum of the squared terms. Using
$h_{\thet}(t_j)\le h_{\thet}(t_r)$ and $\Phi_{\thet}(x_j)\ge1$, each term is bounded
by a constant times $\prod_j\Phi_{\thet}(x_j)$. Thus
	\[
	p_{r,\thet}\|\dot{\boldsymbol\ell}_{r,\thet}\|^2
	\le Ce^{-h_{\thet}(t_r)}(1+h_{\thet}(t_r))^2
	\prod_{j=1}^r\{\lambda_{\thet}(x_j)\Phi_{\thet}(x_j)\}.
	\]
	Partition the ordered $t$-cone according to the number $l$ of
	coordinates below $\delta$. Apply \eqref{eq:joint-negative-small-t} to these coordinates.
	For the other coordinates use $h_{\thet}(t_j)\le h_{\thet}(t_r)$
	and absorb every polynomial factor into the exponential. If $l=r$,
	then $h_{\thet}(t_r)\le1$ and the same bound follows directly.
	After including the Jacobian, the resulting integrand is bounded by
	\[
	C\prod_{j=1}^l
	t_j^{\rho_*-1}(1+|\log t_j|^4)
	\prod_{j=l+1}^r t_j^{k_0-1}.
	\]
	This bound is integrable even without the ordering restriction, since
$\int_0^\delta t^{\rho_*-1}(1+|\log t|^4)\,dt<\infty$ and
$\int_\delta^\infty t^{k_0-1}\,dt=\delta^{k_0}/(-k_0)<\infty$. Since the
gradient vanishes off each candidate support,
	\begin{equation}
	\int_{S_{r,\thetao}}\sup_{\thet\in V}
	\|\nabla s_{\thet}(\mathbf x)\|^2\,d\mathbf x<\infty. \label{eq:joint-negative-gradient-integral}
	\end{equation}
	By \eqref{eq:modified-density-power} with $a=1/2$ and
	\eqref{eq:zero-extension-mvt}, the required envelope on $S_{r,\thetao}$ is
	\[
	\dot m_r^J(\mathbf x)=
	4\frac{\sup_{\thet\in V}\|\nabla s_{\thet}(\mathbf x)\|}
	{p_{r,\thetao}(\mathbf x)^{1/2}}.
	\]
	Its second moment under $P_{\thetao}$ is sixteen times the integral in
\eqref{eq:joint-negative-gradient-integral}.
	This proves the lemma.
\end{proof}

\begin{lemma}[Joint modified criterion at the Gumbel point]
\label{lem:joint-mod-gumbel}
If $k_0=0$, there exist a neighborhood $V$ of $\thetao$ and
$\dot m_r^J\in L^2(P_{\thetao})$ such that
\[
|m^J_{r,\thet_1}-m^J_{r,\thet_2}|
\le \dot m_r^J\|\thet_1-\thet_2\|,
\qquad \thet_1,\thet_2\in V.
\]
\end{lemma}

\begin{proof}
	We adapt the two-tail argument of
\citet[Appendix~D, Case~III]{buchersegers2017maximum} to the joint ordered
density. Put $s_{\thet}=p_{r,\thet}^{1/2}$, with zero extension outside the
support, write $T_{\thet j}=T_{\thet}(x_j)$, and let
	\[
	\lambda_{\thet}(x)=\sigma^{-1}T_{\thet}(x)^{1-k},\qquad
	\Phi_{\thet}(x)=1+\|D_{\thet}(x)\|^2+|\log T_{\thet}(x)|^2,
	\]
	so that $p_{r,\thet}(\mathbf x)=e^{-T_{\thet r}}\prod_{j=1}^r\lambda_{\thet}(x_j)$
	on the support.
	Choose an open convex neighborhood $V$ of
	$\thetao=(\mu_0,\sigma_0,0)$ whose closure has
	$|k|\le\varepsilon$ and $0<c_\sigma\le\sigma\le C_\sigma$,
	where $0<\varepsilon<1/5$ may be decreased below.
By \eqref{eq:D-uniform}, with $t=T_{\thet}(x)$,
	\[
	\|D_{\thet}(x)\|\le
	C(t^{-\varepsilon}+t^\varepsilon)(1+|\log t|^2)
	\]
	uniformly on $V$ and on the candidate support.
The factor $\Phi_{\thet}$ collects the derivative and logarithm terms
that appear in the squared score. For $t\le1$, squaring the bound on $\|D_{\thet}\|$ and multiplying
by $\lambda_{\thet}=\sigma^{-1}t^{1-k}\le Ct^{1-\varepsilon}$ gives
$\lambda_{\thet}\Phi_{\thet}\le Ct^{1-3\varepsilon}(1+|\log t|^4)$, and
$t^{\varepsilon}|\log t|^4\le C_\varepsilon$ turns this into
$Ct^{1-4\varepsilon}$. For $t\ge1$ the same terms are bounded by a power of
$t$. Hence, for a finite $Q$,
	\begin{equation}
	\lambda_{\thet}(x)\Phi_{\thet}(x)\le
	\begin{cases}
	Ct^{1-4\varepsilon},&0<t\le1,\\
	C(1+t)^Q,&t\ge1.
	\end{cases}                                                   \label{eq:joint-gumbel-factor-bound}
	\end{equation}
	All constants here and below are independent of $\thet\in V$.

	By \eqref{eq:joint-score}, the Cauchy--Schwarz inequality and
$T_{\thet j}\le T_{\thet r}$, the squared score is at most
$C(1+T_{\thet r})^2\allowbreak\sum_j\Phi_{\thet}(x_j)$. Since $\Phi_{\thet}\ge1$, the sum
is at most $r\prod_j\Phi_{\thet}(x_j)$, so
	\[
	\|\dot{\boldsymbol\ell}_{r,\thet}(\mathbf x)\|^2
	\le C(1+T_{\thet r})^2\prod_{j=1}^r \Phi_{\thet}(x_j).
	\]
	Thus
	\begin{equation}
	p_{r,\thet}(\mathbf x)\|\dot{\boldsymbol\ell}_{r,\thet}(\mathbf x)\|^2
	\le Ce^{-T_{\thet r}}(1+T_{\thet r})^2
	\prod_{j=1}^r\{\lambda_{\thet}(x_j)\Phi_{\thet}(x_j)\}.       \label{eq:joint-gumbel-product-bound}
	\end{equation}
	
	By Lemma~\ref{lem:sqrt-smooth} (here $|k|\le\varepsilon<1/3$),
	$\thet\mapsto s_{\thet}(\xv)$ is continuously differentiable on $V$ for
	every $\xv$, with $\nabla s_{\thet}=\tfrac12s_{\thet}\dot{\boldsymbol\ell}_{r,\thet}$ on
	the support and $\nabla s_{\thet}=0$ off it; this covers both signs of $k$
	near $k=0$.

	It remains to prove joint domination. Write
	$z_0=(x-\mu_0)/\sigma_0$ and $z_{\thet}=(x-\mu)/\sigma$.
	Choose $M$ large enough that, uniformly in an initial neighborhood of
	$\thetao$, $z_{\thet}\ge cz_0$ for $z_0>M$ and
	$z_{\thet}\le-c|z_0|$ for $z_0<-M$, for some $c>0$.
	Shrink $V$ so that the central region $|z_0|\le M$ stays uniformly
	inside every candidate support. On that region $T_{\thet}$ is bounded
	above and away from zero, and $\lambda_{\thet} \Phi_{\thet}$ is bounded.
	
	For $z_0>M$, $T_{\thet}(x)\le1$ on the candidate support.
	If $k\ge0$, then $T_{\thet}(x)\le e^{-z_{\thet}}$.
	If $k=-\kappa<0$, $0<\kappa\le\varepsilon$, then
	\[
	T_{\thet}(x)=(1+\kappa z_{\thet})^{-1/\kappa}
	\le(1+\varepsilon z_{\thet})^{-1/\varepsilon}.
	\]
	The latter inequality follows because
	$\log(1+\kappa z)/\kappa$ decreases in $\kappa>0$ for $z>0$.
	Combining these estimates with \eqref{eq:joint-gumbel-factor-bound}, and decreasing $\varepsilon$
	so that $\beta=(1-4\varepsilon)/\varepsilon>1$, gives
	\begin{equation}
	\sup_{\thet\in V:\,x\text{ in support}}
	\lambda_{\thet}(x)\Phi_{\thet}(x)
	\le C(1+z_0)^{-\beta}=:\psi_+(z_0),\qquad
	\int_M^\infty \psi_+(z)\,dz<\infty.                         \label{eq:joint-gumbel-upper-envelope}
	\end{equation}
	For $z_0<-M$, put $u=-z_{\thet}>0$.
	For $k>0$, $T_{\thet}=(1+ku)^{1/k}\ge1+u$ when $k<1$;
	for $k=0$, $T_{\thet}=e^u\ge1+u$; and for $k=-\kappa<0$,
	$T_{\thet}=(1-\kappa u)^{-1/\kappa}\ge e^u\ge1+u$ on its support.
	Thus
	\begin{equation}
	T_{\thet}(x)\ge1+c|z_0|                                  \label{eq:joint-gumbel-lower-transform}
	\end{equation}
	uniformly over supported candidates.
	
	Because $x_1>\cdots>x_r$, the coordinates in the upper tail come first and
	those in the lower tail last. Partition the ordered cone accordingly into
	regions $A_{l,b}$, $0\le l\le b\le r$,
	where the first $l$ coordinates satisfy $z_{0j}>M$, the next $b-l$
	satisfy $|z_{0j}|\le M$, and the last $r-b$ satisfy $z_{0j}<-M$.
	On a region with $b<r$, the positive-tail factors in \eqref{eq:joint-gumbel-product-bound} are bounded
	by \eqref{eq:joint-gumbel-upper-envelope}. Every remaining factor is bounded by a finite power of
	$1+T_{\thet r}$, using central boundedness, \eqref{eq:joint-gumbel-factor-bound}, and ordering.
	Since $e^{-t}(1+t)^q\le C_qe^{-t/2}$, \eqref{eq:joint-gumbel-lower-transform} yields
	\[
	\sup_{\thet\in V:\,\mathbf x\text{ in support}}
	p_{r,\thet}(\mathbf x)\|\dot{\boldsymbol\ell}_{r,\thet}(\mathbf x)\|^2
	\le C\prod_{j=1}^l \psi_+(z_{0j})e^{-c|z_{0r}|/2}.
	\]
	For fixed $z_{0r}$ the other negative-tail coordinates occupy intervals
	of total volume at most $C(1+|z_{0r}|)^{r-b-1}$; the central coordinates
	occupy a bounded set. With $y=|z_{0r}|$, its integral is at most
\[
C\left\{\int_M^\infty \psi_+(z)\,dz\right\}^{l}
(2M)^{b-l}\int_M^\infty(1+y)^{r-b-1}e^{-cy/2}\,dy<\infty.
\]
This bound keeps the joint structure of the ordered vector: the
intermediate lower-tail coordinates range over an interval between $z_{0r}$
and $-M$ whose length grows with $|z_{0r}|$, and this growth is absorbed by
the factor $e^{-c|z_{0r}|/2}$. Bounds on the one-dimensional marginals alone
would not control these cross terms.
	If $b=r$ and $l<r$, the last coordinate is central and $T_{\thet r}$
	is uniformly bounded; if $l=b=r$, then $T_{\thet r}\le1$.
	In both cases \eqref{eq:joint-gumbel-product-bound}--\eqref{eq:joint-gumbel-upper-envelope} give an integrable envelope consisting
	of positive-tail factors and bounded central factors. Summing over
	the finite partition proves
	\begin{equation}
	\int\sup_{\thet\in V}
	\|\nabla s_{\thet}(\mathbf x)\|^2\,d\mathbf x<\infty. \label{eq:joint-gumbel-gradient-integral}
	\end{equation}
	(The affine change to $z_{0j}$ contributes only the Jacobian $\sigma_0^r$.)
	
	Finally, \eqref{eq:modified-density-power} with $a=1/2$ and
	\eqref{eq:zero-extension-mvt} give
	\[
	|m^J_{r,\thet_1}(\mathbf x)-m^J_{r,\thet_2}(\mathbf x)|
	\le\dot m_r^J(\mathbf x)\|\thet_1-\thet_2\|,
	\qquad \thet_1,\thet_2\in V,
	\]
	where, on the true support,
	\[
	\dot m_r^J(\mathbf x)=
	4\frac{\sup_{\thet\in V}\|\nabla s_{\thet}(\mathbf x)\|}
	{p_{r,\thetao}(\mathbf x)^{1/2}}.
	\]
	Set this envelope to zero off the true support. By \eqref{eq:joint-gumbel-gradient-integral},
	\[
	\Ep_{\thetao}\{(\dot m_r^J)^2\}
	=16\int_{\{p_{r,\thetao}>0\}}
	\sup_{\thet\in V}\|\nabla s_{\thet}(\mathbf x)\|^2
	\,d\mathbf x<\infty.
	\]
	This proves the lemma.
\end{proof}

\begin{proof}[Proof of Proposition~\ref{prop:rootn}]
	Strong consistency follows from Theorem~\ref{thm:consistency}. All neighborhoods below
	may be chosen within the interior of $K$.
	Write $P_0=P_{\thetao}$ and
	$\mathbb G_n=\sqrt n(P_n-P_0)$.
	For a criterion $m_{\thet}$ with $m_{\thetao}=0$, a local bound
	\begin{equation}
	|m_{\thet_1}-m_{\thet_2}|
	\le\dot m\|\thet_1-\thet_2\|,\qquad P_0\dot m^2<\infty,
	\label{eq:rate-criterion-lipschitz}
	\end{equation}
	implies, in this finite-dimensional parameter space,
	\begin{equation}
	E^*\sup_{\|\thet-\thetao\|\le\delta}
	|\mathbb G_n m_{\thet}|\le C\delta                         \label{eq:rate-empirical-modulus}
	\end{equation}
	for all sufficiently small $\delta$, uniformly in $n$.
	Here $E^*$ denotes outer expectation, which avoids measurability issues
	for the supremum. To see \eqref{eq:rate-empirical-modulus}, consider the
	class $\mathcal M_\delta=\{m_{\thet}:\|\thet-\thetao\|\le\delta\}$. Since
	$m_{\thetao}=0$, \eqref{eq:rate-criterion-lipschitz} shows that
	$\delta\dot m$ is an envelope of $\mathcal M_\delta$, and that two members
	of $\mathcal M_\delta$ are close in $L^2(P_0)$ whenever their parameters are
	close in $\R^3$. More explicitly, an $\epsilon\delta/2$-net
	with centers $\thet_j$ gives brackets
	\[
	\left[m_{\thet_j}-\frac{\epsilon\delta}{2}\dot m,
	      m_{\thet_j}+\frac{\epsilon\delta}{2}\dot m\right].
	\]
	The pointwise bound \eqref{eq:rate-criterion-lipschitz} places each
	nearby criterion in its bracket, whose $L^2(P_0)$-size is
	$\epsilon\delta\|\dot m\|_{P_0,2}$.
	Consequently, $\mathcal M_\delta$ can be covered by
	$N\le C\epsilon^{-3}$ brackets of $L^2(P_0)$-size
	$\epsilon\|\delta\dot m\|_{P_0,2}$, $0<\epsilon<1$
	\citep[Example~19.7]{vanderVaart1998}; the exponent $3$ is the dimension
	of the parameter. The bracketing
	integral $\int_0^1\sqrt{1+\log N}\,d\epsilon$ is therefore finite, and the
	maximal inequality of \citet[Corollary~19.35]{vanderVaart1998} gives
	\eqref{eq:rate-empirical-modulus} with $C$ proportional to
	$\|\dot m\|_{P_0,2}$. The same argument underlies
	\citet[Corollary~5.53]{vanderVaart1998} and
	\citet[Appendix~D]{buchersegers2017maximum}; see also
	\citet[Section~2.14]{vanderVaartWellner1996}.
	
	If $0<k_0<1/2$, use $m_{\thet}=M_{r,\thet}$ from \eqref{eq:sequential-modified}.
	By strong consistency, $\widehat\thet_n$ belongs to this positive-shape
	neighborhood eventually almost surely, so Lemma~\ref{lem:seq-criterion}
	applies for all sufficiently large $n$.
	The conditional factors are defined for every finite conditioning
	value by the positive-shape extension in Lemma~\ref{lem:cond-mod-positive}. In particular,
	\[
	q_{s,\thet}(y\mid x)
	=\frac{f_{\thet}(y)}{F_{\thet}(x)}\mathbf1\{y<x\},\qquad
	\int q_{s,\thet}(y\mid x)\,dy=1,
	\]
	including $x$ above the candidate upper endpoint, where $F_{\thet}(x)=1$.
	Lemma~\ref{lem:seq-criterion} gives $P_n m_{\widehat\thet_n}\ge0$.
	Lemma~\ref{lem:cond-mod-positive} and \citet[Lemma~D.3]{buchersegers2017maximum} for the first
	marginal yield \eqref{eq:rate-criterion-lipschitz}, with
	\[
	\dot m(\mathbf x)=\dot m_1(x_1)
	+\sum_{s=2}^r\dot m_s(x_{s-1},x_s)\in L^2(P_0).
	\]
	The sum is empty when $r=1$.
	The conditional mixture divergences are nonnegative, so Lemma~\ref{lem:seq-drift}
	gives, for some $c>0$,
	\begin{equation}
	P_0m_{\thet}\le P_0m_{1,\thet}
	\le-c\|\thet-\thetao\|^2.                          \label{eq:rate-quadratic-drift}
	\end{equation}
	
	If $k_0\le0$, use $m_{\thet}=m^J_{r,\thet}$.
	The arithmetic--geometric mean inequality and the MLE property give
	\[
	P_nm_{\widehat\thet_n}
	\ge P_n\log\frac{p_{r,\widehat\thet_n}}{p_{r,\thetao}}
	\ge0.
	\]
	Lemma~\ref{lem:joint-mod-negative} when $k_0<0$, and Lemma~\ref{lem:joint-mod-gumbel} when $k_0=0$, yield \eqref{eq:rate-criterion-lipschitz}.
	Let $Q^r_{\thet}=(P^r_{\thet}+P^r_{\thetao})/2$.
	Projection onto $X^{(1)}$ and the data-processing inequality for
relative entropy, a consequence of the chain rule
\citep[Theorem~2.5.3]{CoverThomas2006}, give
	\[
	P_0m^J_{r,\thet}
	=-2D_{\rm KL}(P^r_{\thetao}\|Q^r_{\thet})
	\le-2D_{\rm KL}(P^1_{\thetao}\|Q^1_{\thet})
	=P^1_{\thetao}m_{1,\thet}.
	\]
	Here the projected mixture is exactly
	$Q^1_{\thet}=(P^1_{\thet}+P^1_{\thetao})/2$.
	\citet[Lemma~D.4]{buchersegers2017maximum}, under $\gamma=-k$, gives
	\[
	P^1_{\thetao}m_{1,\thetao+h}
	=-\tfrac14h^TI_1(\thetao)h+o(\|h\|^2).
	\]
	Since $I_1(\thetao)>0$ by Proposition~\ref{prop:info-pd}, \eqref{eq:rate-quadratic-drift} follows in this
	case as well, exactly as in the proof of Lemma~\ref{lem:seq-drift}.
	
	Thus in both cases we have consistency, $P_nm_{\widehat\thet_n}\ge0$,
	the modulus \eqref{eq:rate-empirical-modulus}, and the quadratic drift \eqref{eq:rate-quadratic-drift}.
	The rate now follows from the usual peeling argument
	\citep[Theorem~5.52]{vanderVaart1998}: the event
	$\{\sqrt n\,d_n\ge J\}$ is split into dyadic shells, on each shell the
	quadratic drift forces the empirical process to be large, and the modulus
	\eqref{eq:rate-empirical-modulus} makes this unlikely. We include the short
	argument for completeness.
	Fix a radius $\delta_0>0$ within the neighborhood where these bounds
	hold and put $d_n=\|\widehat\thet_n-\thetao\|$.
	On $0<d_n<\delta_0$, $c d_n^2\le(P_n-P_0)m_{\widehat\thet_n}$, 	because $P_nm_{\widehat\thet_n}\ge0$ and $P_0m_{\widehat\thet_n}\le-cd_n^2$;
	equivalently, $\mathbb G_nm_{\widehat\thet_n}\ge c\sqrt n\,d_n^2$.
	For $j=0,1,\ldots$, on the annulus
$2^jJ/\sqrt n\le d_n<2^{j+1}J/\sqrt n$,
	intersected with $d_n<\delta_0$, the empirical supremum must therefore
	exceed $c2^{2j}J^2/\sqrt n$. Markov's inequality and \eqref{eq:rate-empirical-modulus}, applied
	with radius $\min(2^{j+1}J/\sqrt n,\delta_0)$, bound its probability
	by $C/(2^jJ)$. Summing the finitely many relevant annuli gives
	\[
	\limsup_{n\to\infty}
	P(\sqrt n\,d_n\ge J)\le C/J,
	\]
	because $P(d_n\ge\delta_0)\to0$ by consistency.
	Letting $J\to\infty$ proves
	$\sqrt n(\widehat\thet_n-\thetao)=O_p(1)$.
\end{proof}

\section{Support stability and common-support likelihood bounds}
\label{app:support}

\begin{proof}[Proof of Lemma~\ref{lem:support-loss}]
Write $\omega(\thet)=\mu+\sigma/k$ when $k\ne0$. The argument is the fixed-$r$ version of
\citet[Lemma~E.1]{buchersegers2017maximum}; for a lower endpoint it uses the
Gamma law of $T_r$.

If $0<k_0<1/2$, then for small $\varepsilon$ all nearby shapes are positive and $S_{r,\thet}=\{x_1>\cdots>x_r:x_1<\omega(\thet)\}$. Let $a_\varepsilon=\inf_{\thet\in U_\varepsilon(\thetao)}\omega(\thet)$. Smoothness of $\omega$ gives $\omega_0-a_\varepsilon=O(\varepsilon)$, and hence
\[
P_{\thetao}\{\Xv\notin\overline S_r(\varepsilon)\}
\le P_{\thetao}\{X^{(1)}\ge\omega_0-C\varepsilon\}.
\]
For $d\downarrow0$, $T_{\thetao}(\omega_0-d) =(k_0d/\sigma_0)^{1/k_0}$,
so $P_{\thetao}\{X^{(1)}\ge\omega_0-d\} =O(d^{1/k_0})=o(d^2)$ when $k_0<1/2$.

If $k_0<0$, the support is determined by $x_r>\omega(\thet)$, and
$b_\varepsilon=\sup_{\thet\in U_\varepsilon(\thetao)}\omega(\thet)=\omega_0+O(\varepsilon)$
gives
\[
P_{\thetao}\{\Xv\notin\overline S_r(\varepsilon)\}
\le P_{\thetao}\{X^{(r)}\le\omega_0+C\varepsilon\}.
\]
For $d>0$, $T_{\thetao}(\omega_0+d) =(|k_0|d/\sigma_0)^{-1/|k_0|}$.
Since $T_r\sim\operatorname{Gamma}(r,1)$,
\[
P(T_r\ge y)=e^{-y}\sum_{j=0}^{r-1}\frac{y^j}{j!},
\]
so the support-loss probability is $o(d^m)$ for every fixed $m>0$, in particular
$o(d^2)$.

If $k_0=0$, choose the neighborhood so that $\sigma\ge\sigma_0/2$ and
$|\mu|\le C_\mu$. For $0<k<\varepsilon$, $\omega(\thet)\ge-C_\mu+\sigma_0/(2\varepsilon)$;
for $-\varepsilon<k<0$, $\omega(\thet)\le C_\mu-\sigma_0/(2\varepsilon)$. Both
endpoints thus recede at least at rate $1/\varepsilon$, and for some $c_+,c_->0$,
\[
S_{r,\thetao}\setminus\overline S_r(\varepsilon)
\subset
\{X^{(1)}\ge c_+/\varepsilon\}
\cup
\{X^{(r)}\le-c_-/\varepsilon\}.
\]
The first probability is $O(e^{-c/\varepsilon})$ under the Gumbel upper tail.  For the
second, $T_r\ge C e^{c'/\varepsilon}$ and the Gamma tail is even smaller.  Both are
$o(\varepsilon^2)$.
\end{proof}

\begin{proof}[Proof of Lemma~\ref{lem:loglik-lipschitz}]
	We adapt the observation-dependent admissible sets of
\citet[Lemma~E.2]{buchersegers2017maximum} to the joint support, for either
choice of norm $\|\cdot\|_\nu$, $\nu\in\{\infty,2\}$; balls and common
supports are taken in that norm throughout. Choose $\varepsilon_0$ sufficiently
	small that the closed $2\varepsilon_0$-ball is contained in the
	parameter space and has $\sigma$ bounded away from zero.
	For $\mathbf x\in S_{r,\thetao}$, put
	\[
	\begin{aligned}
	\mathcal A_\nu(\mathbf x)=\bigl\{\thet:\ &
	\|\thet-\thetao\|_\nu<\varepsilon_0,\ \text{and }\mathbf x\in S_{r,\vartheta}\\
	&\text{for every }\vartheta\text{ with }\|\vartheta-\thetao\|_\nu
	\le2\|\thet-\thetao\|_\nu\bigr\}.
	\end{aligned}
	\]
	If $\mathbf x$ is in the common support for the open
	$2\varepsilon$-ball, every parameter in the $\varepsilon$-ball,
	and hence every point of the segment joining two such parameters,
	belongs to $\mathcal A_\nu(\mathbf x)$.
	It suffices to bound the score uniformly on this set by a measurable
	$H\in L^2(P_{\thetao})$. The mean-value theorem then gives the
	asserted inequality with $\dot\ell_r=3H$ for both norms, where $H$
	bounds the Euclidean norm of the score: indeed,
	$|\bm v^T(\thet_1-\thet_2)|\le\|\bm v\|_1\|\thet_1-\thet_2\|_\infty
	\le\sqrt3\,\|\bm v\|_2\|\thet_1-\thet_2\|_\infty$ and
	$\|\thet_1-\thet_2\|_\infty\le\|\thet_1-\thet_2\|_2$.
	
	Suppose first that $0<k_0<1/2$ and write
	$\omega(\thet)=\mu+\sigma/k$, $d=\omega_0-x_1>0$.
	Let $\delta=\|\thet-\thetao\|_\nu$ for
	$\thet\in\mathcal A_\nu(\mathbf x)$.
	Shifting the location by $h$ shifts the endpoint by $h$:
	$\omega(\thet+h\bm e_\mu)=\omega(\thet)+h$. The parameter
	$\thetao-2\delta\bm e_\mu$ belongs to the defining closed ball
	$\{\vartheta:\|\vartheta-\thetao\|_\nu\le2\delta\}$, so $\xv$ lies in its
	support, that is, $x_1<\omega_0-2\delta$; hence $\delta<d/2$. The parameter
	$\thet-\delta\bm e_\mu$ also belongs to that ball, by the triangle
	inequality, so $x_1<\omega(\thet)-\delta$, that is,
	$g=\omega(\thet)-x_1>\delta$. Thus admissibility keeps every fixed
	observation at a distance proportional to $\delta$ from the moving
	endpoint; this is the role of the factor $2$, as in
	\citet[Lemma~E.2]{buchersegers2017maximum}.
	Local Lipschitz continuity of $\omega$ gives
	\[
	|g-d|\le L\delta,\qquad
	\frac{d}{1+L}\le g\le(1+L/2)d.
	\]
	These inequalities also hold when $\delta=0$.
	For every $j$, add $x_1-x_j\ge0$ to both endpoint gaps and use
	$w_{\thet}(x_j)=(k/\sigma)\{\omega(\thet)-x_j\}$.
	Shrinking the neighborhood to keep $k$ positive and bounded away
	from zero yields
	\begin{equation}
	c w_{\thetao}(x_j)\le w_{\thet}(x_j)
	\le C w_{\thetao}(x_j),\qquad j=1,\ldots,r.       \label{eq:common-support-w-comparison}
	\end{equation}
	Put $t_j=T_{\thetao}(x_j)$. Formula \eqref{eq:common-support-w-comparison} and \eqref{eq:D} imply
	\[
	\max_j w_{\thet}(x_j)^{-1}\le Ct_1^{-k_0},\qquad
	T_{\thet}(x_r)\le C(1+t_r^M),
	\]
	and
	\[
	\max_j|\log T_{\thet}(x_j)|
	\le C\{1+|\log t_1|+\log(1+t_r)\}
	\]
	for some finite $M$, uniformly on the admissible set.
	Thus \eqref{eq:joint-score} gives the measurable upper bound
	\begin{equation}
	H(\mathbf x)=C(1+t_r^M)
	\{1+t_1^{-k_0}+|\log t_1|^2+\log^2(1+t_r)\}.           \label{eq:common-support-positive-envelope}
	\end{equation}
	
	If $k_0<0$, put $d=x_r-\omega_0>0$.
	Using $\thetao+2\delta e_\mu$ gives $\delta<d/2$.
Using $\thet+\delta e_\mu$ gives
$g=x_r-\omega(\thet)>\delta$. Again $|g-d|\le L\delta$, so
$d/(1+L)\le g\le(1+L/2)d$.
	Adding $x_j-x_r\ge0$ proves \eqref{eq:common-support-w-comparison} in this case as well.
	Now $\max_j w_{\thet}(x_j)^{-1}\le Ct_r^{-k_0}$, 	and all powers of this quantity can be absorbed into a finite power
	of $1+t_r$. Formulas \eqref{eq:D}--\eqref{eq:joint-score} give, for some finite $M$,
	\begin{equation}
	H(\mathbf x)=C(1+t_r^M)
	\{1+|\log t_1|^2+\log^2(1+t_r)\}.                     \label{eq:common-support-nonpositive-envelope}
	\end{equation}
	
	Finally let $k_0=0$, write
	$z_{0j}=(x_j-\mu_0)/\sigma_0$, and put
	$R=\max(|z_{01}|,|z_{0r}|)$.
	For sufficiently large $R$, if $z_{01}>0$, the parameter
	$(\mu_0,\sigma_0,1/z_{01})$ places its upper endpoint at $x_1$;
	if $z_{0r}<0$, the parameter
	$(\mu_0,\sigma_0,1/z_{0r})$ places its lower endpoint at $x_r$.
	At least one of these endpoints corresponds to $R$.
	Admissibility therefore implies $\delta R<\tfrac12$. 	For $z_j=(x_j-\mu)/\sigma$, $|z_j-z_{0j}|\le C\delta(1+R)\le C$, 	uniformly in the tails. Since $|k|\le\delta$, $|z_{0j}|\le R$ and $\delta R<1/2$,
	we have $|kz_j|\le\delta R+C\delta\le1/2+C\varepsilon_0\le3/4$ after
	decreasing $\varepsilon_0$ if necessary.
	Along the interval between $0$ and $z_j$, we consequently have
	$1-ks\ge1/4$. The integral identities
	\[
	\log T_{\thet}(x_j)=-\int_0^{z_j}\frac{ds}{1-ks},\qquad
	\partial_k\log T_{\thet}(x_j)
	=-\int_0^{z_j}\frac{s\,ds}{(1-ks)^2}
	\]
	imply bounds $4|z_j|$ and $8z_j^2$, respectively.
	The other two derivatives are bounded by $C$ and
	$C(1+|z_j|)$.
	When $z_j\ge0$, $T_{\thet}(x_j)\le1$; when $z_r<0$, $T_{\thet}(x_r)\le e^{4|z_r|}\le C(1+t_r^4), \quad t_r=e^{-z_{0r}}$. 	Here $z_r-z_{0r}$ is bounded; if $z_{0r}\ge0$ but $z_r<0$,
	both are bounded and the same inequality holds after increasing $C$.
	Since $z_{0j}=-\log t_j$, \eqref{eq:joint-score} again yields an envelope of the form
	\eqref{eq:common-support-nonpositive-envelope}. On the remaining bounded-$R$ region, shrink
	$\varepsilon_0$ so that every coordinate stays uniformly inside the
	candidate support; compactness then gives a constant bound.
	
	We finish by checking the squared envelopes, including their mixed
	moments. Under the arrival representation,
	$T_1\sim\operatorname{Exp}(1)$ and $\Delta_r=T_r-T_1$ is independent of
	$T_1$, with $\Delta_r\sim\operatorname{Gamma}(r-1,1)$ for $r\ge2$;
	for $r=1$, set $\Delta_r=0$.
	Using $T_r=T_1+\Delta_r$, logarithmic factors in $1+T_r$ can be absorbed
	into a finite power of $1+T_r$.
	For \eqref{eq:common-support-positive-envelope}, the squared envelope is therefore bounded by $C(1+T_r^q)\{1+T_1^{-2k_0}+|\log T_1|^4\}$ 	for a finite $q>0$. Independence and
	$(T_1+\Delta_r)^q\le C_q(T_1^q+\Delta_r^q)$ reduce its expectation to
	one-dimensional exponential and Gamma moments. The only singular
	power at zero is $T_1^{-2k_0}$, which is integrable because
	$2k_0<1$. The squared envelope \eqref{eq:common-support-nonpositive-envelope} is integrable by the same
	argument without that singular power. Thus $H\in L^2(P_{\thetao})$.
	Setting the envelope to zero outside the true support completes the
	proof for both norms.
\end{proof}

\begin{small}
\renewcommand{\baselinestretch}{1.0}\selectfont
\setlength{\bibsep}{1.5pt plus 0.5pt}
\bibliographystyle{spbasic}
\bibliography{ref_asymp_MLE_15Oct26_Extremes_v7}

\end{small}

\end{document}